\documentclass[12pt]{article}
\usepackage{amsmath}
\usepackage{graphicx,psfrag,epsf}
\usepackage{enumerate}

\newcommand{\blind}{0}

\usepackage{amssymb}
\usepackage{soul}
\usepackage{amsmath}
\usepackage{amsthm}
\usepackage{latexsym}
\usepackage{verbatim}
\usepackage{epsfig}

\usepackage{amsbsy}
\usepackage{amscd}
\usepackage{bm}
\usepackage{graphicx,psfrag,epsf}
\usepackage{enumerate}
\usepackage[authoryear,sort]{natbib}
\setcitestyle{authoryear}
\usepackage{bm}
\usepackage{xcolor}
\definecolor{gmu}{cmyk}{1,0,0.83,0.47}
\usepackage{nicefrac}
\usepackage{url}

\providecommand{\M}[1]{\mathbf#1}
\providecommand{\mc}[1]{\mathcal#1}
\providecommand{\mc}[1]{\mathcal#1}

\newcommand{\R}{{\mathbb R}}

\DeclareMathOperator{\E}{\mathbf{E}}
\DeclareMathOperator{\p}{\mathbf{P}}

\DeclareMathOperator{\cov}{Cov}
\newcommand{\indep}{\rotatebox[origin=c]{90}{$\models$}}

\DeclareMathOperator{\tr}{tr}

\providecommand{\T}{\top} 

\DeclareMathOperator*{\argmin}{argmin}
\DeclareMathOperator*{\argmax}{argmax}

\providecommand{\wt}[1]{\widetilde{#1}}
\providecommand{\wh}[1]{\widehat{#1}}

\newcommand{\dev}[2]{\Big\lvert _{{#1}={#2}}}

\newcommand{\blanco}[1]{  }

\newcommand{\deriv}[3]{%
\ifthenelse{#1 = 1}{\frac{d\,#2}{d\,#3}}{\frac{d^{{#1}} #2}{d{#3}^{{#1}}}}
}

\newcommand{\partials}[3]{%
\ifthenelse{#1 = 1}{\frac{\partial\,#2}{\partial\,#3}}{\frac{\partial^{#1}
    #2}{\partial#3^{#1}}}
} 

\def\su{\sum_{i=1}^n}

\def \coloneq{\mathrel{\mathop:}=}

\def \eps{\varepsilon}

\newcommand{\tcb}[1]{\textcolor{black}{#1}}

\begin{document}

\def\spacingset#1{\renewcommand{\baselinestretch}%
{#1}\small\normalsize} \spacingset{1}


\if0\blind
{
  \title{\bf A General Framework for Regression with Mismatched Data Based on Mixture Modeling}
  \author{\begin{tabular}{lll} Martin Slawski$^{1}$\thanks{
    Corresponding Author; Partially supported by NSF grants \#1849876 and \#2120318} & $\quad$ Brady T. West$^{2\dagger}$ & $\quad$ Priyanjali Bukke$^{1}\thanks{Partially supported by NSF grant \#2120318}$ \\ Zhenbang Wang$^{1\dagger}$ & $\quad$ Guoqing Diao$^{3}$ & $\quad$ Emanuel Ben-David$^{4}$ \end{tabular} \\[3ex]
    \begin{tabular}{l}
    $^{1}${\normalsize Department of Statistics, George Mason University}\\
    $^{2}${\normalsize Institute for Social Research, University of Michigan}\\
    $^{3}${\normalsize Department of Biostatistics and Bioinformatics, George Washington University}\\
    $^{4}${\normalsize Center for Statistical Research \& Methodology, U.S.~Census Bureau}\\
    \end{tabular}
   }
  \maketitle

} \fi

\if1\blind
{
  \bigskip
  \bigskip
  \bigskip
  \begin{center}
    {\LARGE\bf Title}
\end{center}
  \medskip
} \fi
\vspace*{-2ex}
\bigskip
\begin{abstract}
Data sets obtained from linking multiple files are frequently affected by mismatch error, as a result of non-unique or noisy identifiers used during record linkage. Accounting for such mismatch error in downstream analysis performed on the linked file is critical to ensure valid statistical inference. In this paper, we present a general framework to enable valid post-linkage inference in the challenging secondary analysis setting in which only the linked file is given. The proposed framework covers a wide selection of statistical models and can flexibly incorporate additional information about the underlying record linkage process. Specifically, we propose a mixture model for pairs of linked records whose two components reflect distributions conditional on match status, i.e., correct match or mismatch. Regarding inference, we develop a method based on composite likelihood and the EM algorithm as well as an extension towards a fully Bayesian approach. Extensive simulations and several case studies involving contemporary record linkage applications corroborate the effectiveness of our framework.    
\end{abstract}

\noindent%

\spacingset{1.45}
\section{Introduction}\label{sec:intro}
Record linkage (RL) \citep[e.g.][]{Newcombe, Binette2022, Christen2012} is an important method of data integration. RL refers to micro-level (i.e., record-by-record) combination of data from multiple sources and thus can be considered the most granular among a variety of approaches to data integration \citep{Lohr2017}. RL comes with the great promise of creating richer data sets from existing ones at virtually no additional cost. Examples include linkage of surveys and administrative records, insurance claims and hospital records, birth and death registries, historical censuses, among many others.    

Harnessing the opportunities associated with RL is not always straightforward since the process of identifying matching records can be error-prone. For instance, privacy considerations often prevent the use of personal identifying information for the purpose of RL. Missing data and quality issues (e.g., formatting or spelling variations) can induce substantial uncertainty, with one record yielding many candidate matches in the other file. Probabilistic RL techniques, e.g. those based on the Fellegi-Sunter method \citep{Fellegi69} address such uncertainty systematically by assigning a matching score to each pair of records, but by no means guarantee that the resulting linked file is free of errors. Mainstream implementations of probabilistic RL can be sensitive to the choice of the threshold for the matching score at which a pair is deemed a match. Proper choice of this threshold strikes a suitable balance between false matches (henceforth {\em mismatches}) and false non-matches ({\em missed matches}). Both types of errors can negatively affect 
downstream statistical analyses ({\em post-linkage analysis}) performed on the linked file. 

While missed matches can induce sample selection bias similar to non-response in survey data \citep{LittleRubin2019}, mismatches can cause data contamination and typically attenuated relationships when analyzing associations, e.g., in regression analysis. This is a well-studied problem dating back to \cite{Neter65}, and important follow up work was conducted by \cite{Scheuren93, Scheuren97} and \cite{Lahiri05}. Subsequently, a variety of approaches have been proposed to account for mismatches in post-linkage data analysis. This body of work can be roughly divided according to whether it addresses {\em primary analysis} or {\em secondary analysis}. The former refers to scenarios in which record linkage and downstream analysis is performed by the same individual, or the data analyst has at least significant insights into the details of the underlying RL. In these situations, it is possible to directly propagate the uncertainty from RL; examples of specific approaches include \cite{Han2019, Hof2014} and various hierarchical Bayes methods \citep[e.g.][]{Gutman13, Dalzell2018, Tancredi15, Steorts2018}. 

By contrast, in the secondary analysis setting, the data analyst only has access to the linked file rather than the individual files and only has limited knowledge about how RL was performed. For instance, the data analyst may be given scores reflecting the likelihood of every linked record being a correct match as in the recent study by \cite{Abowd2019}, or indicators of the blocks within which linkage was performed as well as the mismatch rate within each block. A line of research pioneered by \cite{Chambers2009} hinges on these pieces of information, typically in conjunction with the assumption of exchangeable linkage error (ELE) within each block. Significant follow-up work along this line includes \citep{KimChambers2012, Zhang2020, Chambers2019improved}. We also refer to two recent survey papers \citep{Wang2022, Chambers2023} and the references therein. 
\vskip1ex
\noindent {\bfseries Contributions}. In this paper, we develop a general framework to account for mismatch error in post-linkage analysis in the secondary analysis setting. This framework generally does not require any information from RL, and provides estimates of mismatch rates in an integrated fashion. At the same time, our method can easily incorporate such information (e.g., block indicators or quantities informative of match status) if available. Moreover, various forms 
of post-linkage analysis can be accommodated under a common umbrella, in particular various forms of regression modeling as well
as covariance estimation and the analysis of contingency tables; specific instances are highlighted in $\S$\ref{sec:examples} below. In a nutshell, the proposed framework relies on a two-component mixture model whose components reflect the latent match status (correctly or incorrectly matched) for each record in the linked file. Estimation is based on composite likelihood \citep{Lindsay1988, Varin2011}, which provides a path towards valid (asymptotic) inference; we also sketch how the proposed method can be cast in a Bayesian framework. The proposed approach extends prior work \citep{SlawskiDiaoBenDavid2019} motivated by ``shuffled data problems"  \citep{DeGroot1980, Pananjady2016, SlawskiBenDavid2017} in multiple directions. In brief, the paper by \cite{SlawskiDiaoBenDavid2019} is limited to classical linear regression and a constant mismatch rate. The approach presented herein bears a close connection to the method in   
\cite{Hof2014}. The main distinction is that the latter method is developed for the primary analysis setting and involves a pair-wise composite likelihood, which renders the approach less scalable. Apart from that, we employ additional assumptions; while these assumptions may be considered strong, they render inference much more tractable.

\vskip1ex
\noindent {\bfseries Organization}. Formal descriptions of the setup, our approach, and the underlying assumptions are provided in $\S$\ref{sec:meth}. We then outline the framework for inference in $\S$\ref{sec:inference}. Specific examples of interest are discussed in more detail in $\S$\ref{sec:examples}. Additional technical details and extensions are presented in $\S$\ref{sec:misc}. Simulation studies and real data analysis are presented in $\S$\ref{sec:simulations} and $\S$\ref{sec:apps}, respectively. We conclude with a summary of the main findings and discuss potential directions for future work in $\S$\ref{sec:conc}.      
\vskip1ex
\noindent {\bfseries Notation}. Here, we summarize notation used repeatedly in subsequent sections of this paper. We use the following conventions regarding probability density functions (PDFs): instead of writing 
$f_{\M{x}}(\M{x}_0)$ for the density of a random vector $\M{x}$ evaluated at a point $\M{x}_0$, we drop the symbol in the subscript and simply write $f(\M{x}_0)$ with the convention that the corresponding random variable is inferred from the symbol in the argument. Similar conventions are adopted for joint and conditional PDFs, i.e, we use $f(\M{a}_0, \ldots, \M{z}_0)$ instead of $f_{\M{a} \ldots \M{z}}(\M{a}_0, \ldots, \M{z}_0)$ and $f(\M{x}_0|\M{y}_0)$ instead of $f_{\M{x}|\M{y} = \M{y}_0}(\M{x}_0)$ etc. Note that subscripts in $f$ will be present in case there is no argument. By default, symbols will be boldfaced to indicate vector-valued quantities, with the understanding that boldfaced quantities may also represent scalars as special case; occasionally, normal instead of bold font is used to highlight a scalar quantity. Finally, dependence of PDFs on parameters is expressed via $f(\cdot\,;\ldots)$, where $\ldots$ represents a list of parameters.  A table summarizing frequently used symbols and notation is given below. \\[3ex]
\begin{tabular}{||l|l||l|l|}
\hline
$\mathbb{I}(\cdot)$  & indicator function  & $\M{u} \indep \M{v}$ & random variables $\M{u}$ and $\M{v}$ are independent \\ 
$m$ & mismatch indicator & $\phi(\M{y}|\M{x})$  & conditional PDF of $\M{y}$ given $\M{x}$ (regression setup) \\
$\M{P}(\ldots)$    & probability                    & $\bm\theta$ & parameter describing the $(\M{x},\M{y})$-relationship \\
$\M{E}[\ldots]$    & expectation & $\M{z}$ & covariates informative of mismatch indicator \\      
$[\ldots]^{(t)}$  & iteration counter & $h(\M{z})$ & $\p(m = 0 | \M{z})$\\
$\text{logit}(x)$  & $\log(x/(1-x))$ & $\bm{\gamma}$ & parameter associated with $h$ \\   
    & & $\bm{\theta}^{\ast}, \bm{\gamma}^{\ast}$ \text{etc}. & ``ground truth" parameter values \\ 
\hline    
\end{tabular}



\section{Methods}\label{sec:meth}
The goal of record linkage is to merge two files individual files $F_{\M{x}}^{\star} = \{ \M{x}_{j}^{\star} \}_{j = 1}^M$ and  $F_{\M{y}}^{\star} = \{ \M{y}_k ^{\star}\}_{k = 1}^N$ into a new 
file $F_{\M{x}\Join\M{y}}^{\star} = \{ (\M{x}_{\ell_i}^{\star}, \M{y}_{\ell_i}^{\star}) \}_{i = 1}^{\nu}$ of pairs corresponding to identical statistical units. For simplicity, we assume that every
$\M{y}_i^{\star}$, $1 \leq i \leq N$, has one and only one match in $F_{\M{x}}^{\star}$ (and hence $M \geq N = \nu$). We also assume that the missing links in the larger file 
$F_{\M{x}}^{\star}$ are ignorable\footnote{Missing at random in  regression settings with the $\M{y}$ variable as the response, missing completely at random in unsupervised settings. See $\S$\ref{subsec:up} for a definition of ``unsupervised settings".}. Data linkage is assumed to produce an imperfectly combined file $F_{\M{x} \Join \M{y}} = \{(\M{x}_i, \M{y}_i) \}_{ i =1}^n$ with $\M{x}_i \in F_{\M{x}}^{\star}$ and $\M{y}_i \in F_{\M{y}}^{\star}$, $1 \leq i \leq n \leq N$, containing \emph{mismatched pairs} $(\M{x}_i, \M{y}_i) \notin F_{\M{x}\Join\M{y}}^{\star}$ and lacking correct matches $\M{F}_{\M{x} \Join \M{y}}^{\star} \setminus \M{F}_{\M{x} \Join \M{y}}$ (\emph{missed matches}). Throughout this paper, we focus on mismatches and assume that missed matches are ignorable. 

With each linked pair in $\M{F}_{\M{x} \Join \M{y}}$, we may additionally observe variables $\M{z}_i$ pertaining to the confidence in the correctness of the link, $1 \leq i \leq n$, which yields triplets $\{ (\M{x}_i, \M{y}_i, \M{z}_i) \}_{i = 1}^n$. Accordingly, we define 
latent mismatch indicators $m_i = \mathbb{I}( (\M{x}_i, \M{y}_i) \notin \M{F}_{\M{x}\Join\M{y}}^{\star})$, $1 \leq i \leq n$. 
\vskip1ex
\noindent \emph{Assumptions}. 
\begin{itemize}
\item[{\bfseries (A1)}] The $\{ (m_i, \M{z}_i) \}_{i = 1}^n$ are independent of both $\{ \M{x}_i \}_{i = 1}^n$ and $\{ \M{y}_i \}_{i = 1}^n$. 
\item[{\bfseries (A2)}] The following two-component mixture model is assumed for each pair $(\M{x}_i, \M{y}_i)$: 
\begin{equation}\label{eq:mixture_model}
(\M{x}_i, \M{y}_i)|\M{z}_i, \{ m_i = 0  \} \sim \phi_i(\cdot;\bm{\theta}^*), \qquad \qquad \qquad \text{(IND):}\; \, \M{y}_i \indep \,\M{x}_i|\M{z}_i, \{ m_i = 1 \},
\end{equation}
where the $\phi_i(\cdot;
\bm{\theta}^*)$ are probability density functions (PDFs) depending on an unknown parameter of interest $\bm{\theta}^*$ but neither on $m_i$ nor on $\M{z}_i$, $1 \leq i \leq n$. The second item in \eqref{eq:mixture_model} will be referred to via the abbreviation 
(IND) in the sequel. 
\item[{\bfseries (A3)}]  $\p(m_i = 0 | \M{z}_i) = h(\M{z}_i;\bm{\gamma}^*)$ for some known function $h$ and 
and unknown parameter $\bm{\gamma}^*$ (of secondary interest), $1 \leq i \leq n$. 
\end{itemize}
\tcb{The above assumptions are satisfied in typical probabilistic record linkage setups as long
(i) $\M{x}$'s and $\M{y}$'s are independent for non-matching records, i.e., $\M{x}_j^{\star} \indep \M{y}_k^{\star}$ for all pairs contained in $\M{F}_{\M{x}}^{\star} \times \M{F}_{\M{y}}^{\star} \setminus \M{F}_{\M{x} \Join \M{y}}^{\star}$, and (ii) record linkage does not depend on the data to be linked themselves. For example, suppose that each candidate pair $(\M{x}_j^{\star}, \M{y}_k^{\star})$ is assigned a comparison vector $\M{c}_{jk}$ based on several quasi-identifiers available in both files not depending on $(\M{x}_j^{\star}, \M{y}_k^{\star})$, $1 \leq j \leq M$, $1 \leq k \leq N$. Suppose further that $(\M{x}_j^{\star}, \M{y}_k^{\star})$ is declared a match (and hence to be included in $\M{F}_{\M{x} \Join \M{y}}$) if $z_{jk} \coloneq \M{c}_{jk}^{\T} \M{w} \geq \tau$ for some fixed weight vector $\M{w}$ and a threshold $\tau$, $1 \leq j \leq M$, $1 \leq k \leq N$, and that the resulting triplets are given by
\begin{equation*}
(\M{x}_i, \M{y}_i, \M{z}_i) = (\M{x}_{j_i}^{\star}, \M{y}_{k_i}^{\star}, z_{j_i \, k_i}), \qquad 1 \leq i \leq n, 
\end{equation*}
where $n = |\{ (j,k): \;  z_{jk} \geq \tau \}|$. In this scenario, inclusion in $\M{F}_{\M{x} \Join \M{y}}$ does not depend on $\M{F}_{\M{x}}^{\star}, \M{F}_{\M{y}}^{\star}$, and the match
status $m_i$ of each included pair only depends on the observed data via $\M{z}_i$, $1 \leq i \leq n$. In general, however, the variables used for linkage may exhibit correlations with the variables to be analyzed after linkage. Moreover, the independence assumption (IND) in {\bfseries (A2)} may be violated, e.g., when mismatches occur among correlated $(\M{x}, \M{y})$-records; a typical example is the often-induced correlation by blocking variables used for record linkage. 
}   

\begin{figure}
\begin{center}
\includegraphics[height = 0.13\textheight]{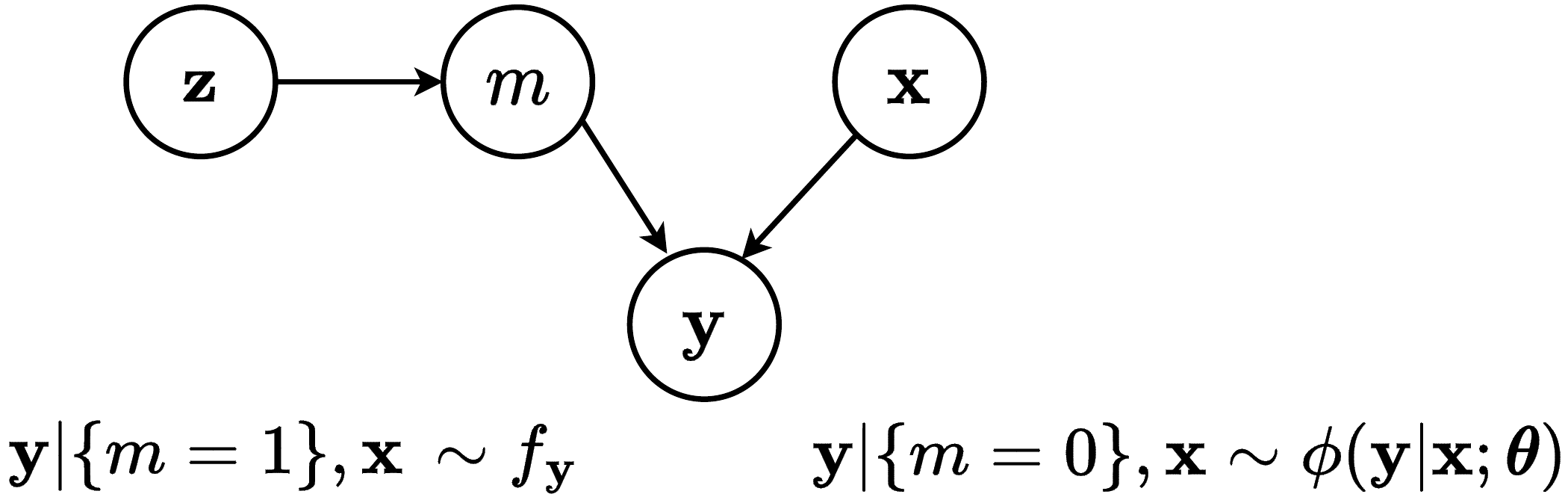}
\end{center}
\vspace*{-2.5ex}
\caption{DAG representation of the model associated with the likelihood \eqref{eq:mixture_likelihood_regression} for regression settings. Note that the covariates for modeling the latent match indicator and the covariates for modeling the response variable are assumed to be independent.}\label{fig:DAG}
\end{figure}

\vskip2ex
\noindent Given {\bfseries (A1)} through {\bfseries (A3)}, the likelihood in the parameters $(\bm{\theta}, \bm{\gamma})$ of a single triplet $(\M{x}_i, \M{y}_i, \M{z}_i)$ is given by
\begin{align}\label{eq:mixture_likelihood}
L_i(\bm{\theta}, \bm{\gamma}) &= f(\M{x}_i, \M{y}_i, \M{z}_i; \bm{\theta}, \bm{\gamma}) \notag \\
                              &= \sum_{m_i \in \{0,1\}} f(\M{x}_i, \M{y}_i, \M{z}_i, m_i; \bm{\theta}, \bm{\gamma}) \notag \\
                              &=\sum_{m_i \in \{0,1\}} f(\M{x}_i, \M{y}_i | \M{z}_i, m_i; \bm{\theta}, \bm{\gamma}) \, f(\M{z}_i, m_i; \bm{\gamma}) \notag \\
                              &\overset{\text{{\bfseries (A2)}}}{=} f(\M{x}_i| \M{z}_i, \{ m_i = 1\}) \times f(\M{y}_i| \M{z}_i, \{ m_i = 1\}) \times f(\M{z}_i, \{ m_i = 1 \}; \bm{\gamma}) + \notag \\
                              &\quad + \phi_i(\M{x}_i, \M{y}_i; \bm{\theta}) \times f(\M{z}_i, \{ m_i = 0 \}; \bm{\gamma}) \notag \\
                              &\overset{\text{{\bfseries (A1)}}}{=} f(\M{x}_i) \times f(\M{y}_i) \times \p(m_i = 1 | \M{z}_i; \bm{\gamma}) \times f(\M{z}_i) +  \phi_i(\M{x}_i, \M{y}_i; \bm{\theta}) \p(m_i = 0 | \M{z}_i; \bm{\gamma}) \times f(\M{z}_i) \notag \\
                              &\overset{\text{{\bfseries (A3)}}}{\propto} f(\M{x}_i;\bm{\theta}) \times f(\M{y}_i;\bm{\theta}) \times \left\{ 1 - h(\M{z}_i;\bm{\gamma}) \right\}  + \phi_i(\M{x}_i, \M{y}_i; \bm{\theta}) \times h(\M{z}_i;\bm{\gamma}), 
\end{align}
where $\propto$ here means equality up to multiplicative constants not involving $\bm{\theta}$ or $\bm{\gamma}$. 
\vskip1ex
\noindent In a regression setup, $\phi_i(\cdot;\bm{\theta})$ depends on $\bm{\theta}$ only via the conditional PDF of the response variable 
given covariates, here denoted by $\phi_i(\cdot | \, \cdot \,; \bm{\theta})$, $1 \leq i \leq n$. The likelihood \eqref{eq:mixture_likelihood} accordingly can be decomposed as 
\begin{align}\label{eq:mixture_likelihood_regression}
L_i(\bm{\theta}, \bm{\gamma}) &= f(\M{x}_i;\bm{\theta}) \times f(\M{y}_i;\bm{\theta}) \times \left\{ 1 - h(\M{z}_i;\bm{\gamma}) \right\} +  \phi_i(\M{y}_i|\M{x}_i;\bm{\theta}) \times f(\M{x}_i; \bm{\theta}) \times  h(\M{z}_i;\bm{\gamma}) \notag \\
&\propto f(\M{y}_i;\bm{\theta}) \times \left\{ 1 - h(\M{z}_i;\bm{\gamma}) \right\} +  \phi_i(\M{y}_i|\M{x}_i;\bm{\theta}) \times  h(\M{z}_i;\bm{\gamma}),\quad 1 \leq i \leq n. 
\end{align}
The corresponding DAG representation is shown in Figure \ref{fig:DAG}. Observe that as a consequence
of {\bfseries (A1)}, the latent match indicator only depends on the $\{ \M{z}_i \}_{i = 1}^n$ but
not on the covariates $\{ \M{x}_i \}_{i = 1}^n$. 

We here briefly note that while the marginal densities $f(\M{x})$ and $f(\M{y})$ are typically not known, their estimation is not affected by mismatch error and thus straightforward; we refer to $\S$\ref{subsec:estimation_marginals} for more details.

\vskip1ex
\noindent \emph{Pseudo-likelihood}. Multiplication of the individual terms $L_i(\bm{\theta}, \bm{\gamma})$, $1 \leq i \leq n$, yields the pseudo likelihood (also known as the composite likelihood)
\begin{equation}\label{eq:PL}
L(\bm{\theta}, \bm{\gamma}) = \prod_{i = 1}^n L_i(\bm{\theta}, \bm{\gamma}),
\end{equation}
to be maximized with respect to $\bm{\theta}$ and $\bm{\gamma}$. Note that the product of the observation-specific $L_i$'s (referred to as ``marginal likelihoods") does not coincide with the joint likelihood associated with the entire
collection of pairs $\{ (\M{x}_i, \M{y}_i) \}_{i = 1}^n$ since the mismatch indicators $\{ m_i \}_{i = 1}^n$ are not independent in general\footnote{This can be verified for simple examples: consider $n = 2$ and suppose that the underlying matching pairs 
are swapped with a certain probability.}, hence the term pseudo-likelihood. Regardless of that, its maximizers can be shown 
to be $\sqrt{n}-$consistent and asymptotic normal with covariance matrix of a sandwich form \cite[e.g.,][]{Lindsay1988, Varin2011}; see $\S$\ref{subsec:standarderrors} below for further details. 

\vskip2ex
\noindent \emph{Secondary Analysis setting}. We here emphasize that the proposed approach is motivated by
secondary analysis in which no additional information beyond the imperfectly linked file $F_{\M{x} \Join \M{y}}$ may be available. In particular, none of the two individual files $F_{\M{x}}^{\star}, F_{\M{y}}^{\star}$ may be given. Additional information from the linkage process such as match probabilities can be incorporated in terms of the variables $\{ \M{z}_i \}_{i = 1}^n$ in a model for the mismatch indicators (cf.~$\S$\ref{subsec:mismatch_modeling} below). Note that 
the $\{ \M{z}_i \}_{i = 1}^n$ are allowed to be empty, in which case the $\{ m_i \}_{i = 1}^n$ are treated as identically distributed Bernoulli random variables. \vskip2ex
\noindent 

A related approach addressing the \emph{primary analysis} setting (in which linkage and subsequent data analysis are considered in an integrated fashion) is developed in \cite{Hof2014}. Their formulation
is based on a \emph{pairwise} pseudo-likelihood over all pairs $F_{\M{x}}^{\star} \times F_{\M{y}}^{\star}$ and associated comparison vectors $\{ \M{c}_{jk}\}$. Apart from that 
the assumptions underlying the specific pseudo-likelihood in \cite{Hof2014} are different, notably in 
that an analog to our assumption {\bfseries (A1)} is not employed, which in turn prompts a different
path to inference. 

\section{Inference}\label{sec:inference}
In the following, we describe the main ingredients of our inferential framework, with specific details and extensions postponed to $\S$\ref{sec:misc}. Selected examples are reviewed in $\S$\ref{sec:examples}. 

\subsection{EM Algorithm}\label{subsec:optimization}
Direct maximization of the pseudo-likelihood tends to be challenging. Treating the mismatch indicators $\{ m_i \}_{i = 1}^n$ as missing data naturally prompts the use of the EM algorithm. The resulting E-step involves simple closed form updates akin to those in conventional mixture models, and the M-step updates for $\bm{\theta}$ and $\bm{\gamma}$ decouple into separate optimization problems. Moreover, the update for $\bm{\theta}$ typically reduces to an optimization problem that would be encountered in the \emph{absence of mismatches} with additional observation weights. As a result, existing software can 
be used as long as these weights can be incorporated. The general template is presented below, assuming for now that 
$f_{\M{x}}$ and $f_{\M{y}}$ are known; we refer to $\S$\ref{subsec:estimation_marginals} for details on this aspect.

The complete data (pseudo-)likelihood is given by 
\begin{align*}
L^{\textsf{c}}(\bm{\theta}, \bm{\gamma}) &= \prod_{i = 1}^n f(\M{x}_i, \M{y}_i, \M{z}_i, m_i; \bm{\theta}, \bm{\gamma}) \\
&\propto \prod_{i = 1}^n f(\M{x}_i, \M{y}_i | \M{z}_i, m_i; \bm{\theta}, \bm{\gamma}) \, f(m_i|\M{z}_i; \bm{\gamma}) \\
&= \prod_{i = 1}^n \big\{ [ f(\M{x}_i) \times f(\M{y}_i) \times (1 - h(\M{z}_i;\bm{\gamma})) ]^{m_i} \times \big [\phi_i(\M{x}_i, \M{y}_i; \bm{\theta}) \times h(\M{z}_i;\bm{\gamma})]^{1 - m_i} \big\},
\end{align*}
where for the last line we use the assumption that  $f_{\M{x}}$ and $f_{\M{y}}$ known. Taking logarithms, the complete data negative (pseudo)-log-likelihood is given by (modulo additive constants): 
\begin{align}
\ell^{\textsf{c}}(\bm{\theta}, \bm{\gamma}) &= -\sum_{i = 1}^n \left\{ m_i \log(1 - h(\M{z}_i;\bm{\gamma})) + (1 - m_i) \log(h(\M{z}_i;\bm{\gamma})) \right\} - \notag\\
&\quad -\sum_{i = 1}^n (1-m_i) \log(\phi_i(\M{x}_i, \M{y}_i;\bm{\theta})) \label{eq:log_complete}
\end{align}

{\bfseries E-step}. In the E-step, we evaluate the conditional expectation of the indicator variables $\{ \mathbb{I}(m_i = 0) \}_{i = 1}^n$ given $\{ (\M{x}_i, \M{y}_i, \M{z}_i) \}_{i = 1}^n$ and current iterates $(\bm{\theta}^{(t)}, \bm{\gamma}^{(t)})$ for the parameters as follows. 
\begin{align}\label{eq:Estep_general}
\p(m_i = 0 | (\M{x}_i, \M{y}_i, \M{z}_i);\bm{\theta}^{(t)},\bm{\gamma}^{(t)}) &= \frac{f(\M{x}_i, \M{y}_i, \M{z}_i | m_i = 0;\bm{\theta}^{(t)},\bm{\gamma}^{(t)}) \times \p(m_i = 0;\bm{\theta}^{(t)},\bm{\gamma}^{(t)})}{\sum_{m \in \{0,1\}}f(\M{x}_i, \M{y}_i, \M{z}_i | m_i = m;\bm{\theta}^{(t)},\bm{\gamma}^{(t)}) \times \p(m_i = m;\bm{\theta}^{(t)},\bm{\gamma}^{(t)})} \notag \\
&=\frac{f(\M{x}_i, \M{y}_i | \{ m_i = 0 \}, \M{z}_i;\bm{\theta}^{(t)},\bm{\gamma}^{(t)}) \times \p(m_i = 0 | \M{z}_i;\bm{\gamma}^{(t)})}{\sum_{m \in \{0,1\}} f(\M{x}_i, \M{y}_i | \{ m_i = m \}, \M{z}_i;\bm{\theta}^{(t)},\bm{\gamma}^{(t)}) \times \p(m_i = m | \M{z}_i;\bm{\gamma}^{(t)})} \notag \\
&= \frac{\phi_i(\M{x}_i, \M{y}_i ; \bm{\theta}^{(t)}) \times h(\M{z}_i;\bm{\gamma}^{(t)})}{\phi_i(\M{x}_i, \M{y}_i ; \bm{\theta}^{(t)}) \times h(\M{z}_i;\bm{\gamma}^{(t)}) + f(\M{x}_i) \times f(\M{y}_i) \times (1 - h(\M{z}_i;\bm{\gamma}^{(t)}))},
\end{align}
$1 \leq i \leq n$, where the final result is obtained by invoking Assumptions {\bfseries (A1)} through {\bfseries (A3)} in $\S$\ref{sec:meth}. 
\vskip1ex
\noindent {\bfseries E-step} (Regression case). Observe that in regression setups with the $\{ \M{x}_i \}_{i = 1}^n$ being conditioned on, the E-step can further be simplified as 
\begin{equation}\label{eq:Estep_regression}
\p(m_i = 0 | (\M{x}_i, \M{y}_i, \M{z}_i);\bm{\theta}^{(t)},\bm{\gamma}^{(t)}) =\frac{\phi_i(\M{y}_i|\M{x}_i ; \bm{\theta}^{(t)}) \times h(\M{z}_i;\bm{\gamma}^{(t)})}{\phi_i(\M{y}_i|\M{x}_i ; \bm{\theta}^{(t)}) \times h(\M{z}_i;\bm{\gamma}^{(t)}) + f(\M{y}_i) \times (1 - h(\M{z}_i;\bm{\gamma}^{(t)}))}, \; 1 \leq i \leq n.
\end{equation}

{\bfseries M-step}. Let $\wh{m}_i^{(t)} = \p(m_i = 1| (\M{x}_i, \M{y}_i, \M{z}_i);\bm{\theta}^{(t)},\bm{\gamma}^{(t)})$, $1 \leq i \leq n$. The \emph{expected} complete data negative (pseudo)-log likelihood then takes the form
\begin{align}
\ell^{(t)}(\bm{\theta}, \bm{\gamma}) &= -\sum_{i = 1}^n \left\{ \wh{m}_i^{(t)} \log(1 - h(\M{z}_i;\bm{\gamma})) + (1 - \wh{m}_i^{(t)}) \log(h(\M{z}_i;\bm{\gamma})) \right\} - \notag\\
&\quad -\sum_{i = 1}^n (1 - \wh{m}_i^{(t)}) \log(\phi_i(\M{x}_i, \M{y}_i;\bm{\theta})) \label{eq:Mstep}
\end{align}
Note that minimization over $\bm{\gamma}$ involves only the first term, which is seen to be the log-likelihood of 
a binary regression model with ``responses" $\{ \wh{m}_i^{(t)} \}_{i = 1}^n$, covariates $\{ \M{z}_i \}_{i = 1}^n$, and link function $h$. Minimization over $\bm{\theta}$ involves the log likelihood encountered in the absence of mismatches with additional observation-specific weights.

\subsection{Standard errors}\label{subsec:standarderrors}
For fully parametric models, asymptotic standard errors can be obtained from well-known properties of composite maximum likelihood estimators. Specifically, letting $(\wh{\bm{\theta}}_n, \wh{\bm{\gamma}}_n)$ denote the maximizer of the pseudo-likelihood \eqref{eq:PL} and $(\bm{\theta}^*, \bm{\gamma}^*)$ the corresponding population parameters, we have (under suitable regularity conditions) that 
\begin{equation*}
\sqrt{n} \left( \begin{array}{c}
\wh{\bm{\theta}}_n \\ 
\wh{\bm{\gamma}}_n
\end{array} \right) \rightarrow  N \left( \left( \begin{array}{c}
\bm{\theta}^* \\ 
\bm{\gamma}^*
\end{array} \right),  \, \E[\nabla^2 \ell(\bm{\theta}^*, \bm{\gamma}^*)]^{-1}\, \E[\nabla \ell(\bm{\theta}^*, \bm{\gamma}^*)^{\otimes2}]  \,  \E[\nabla^2 \ell(\bm{\theta}^*, \bm{\gamma}^*)]^{-1}   \right),
\end{equation*}
in distribution, where $\nabla \ell$ and $\nabla^2 \ell$ denote the gradient and Hessian of $\ell = -\log L$ and $\M{v}^{\otimes 2} = \M{v}\M{v}^{\T}$ denotes the outer product of a vector $\M{v}$. Moreover, the above covariance can be estimated consistently by substituting the expectations with their corresponding empirical averages evaluated at  $(\wh{\bm{\theta}}_n, \wh{\bm{\gamma}}_n)$, i.e.,  
\begin{equation}\label{eq:score_hessian}
\frac{1}{n} \su \nabla \ell_i(\wh{\bm{\theta}}_n, \wh{\bm{\gamma}}_n)^{\otimes2}, \qquad \frac{1}{n} \su \nabla^2 \ell_i(\wh{\bm{\theta}}_n, \wh{\bm{\gamma}}_n),
\end{equation}
where $\ell_i$ represents the $i$-th summand in $\ell$, i.e., $\ell = \su \ell_i$. More specific expressions for the above quantities are provided in the appendix. 


\section{Specific examples}\label{sec:examples}

In this section, we work out of the specifics of the general template in the previous section for several popular
regression setups. We also present applications to covariance estimation and contingency table analysis. Modeling 
of the latent mismatch indicators is discussed in a dedicated subsection. 

\subsection{Linear Regression}\label{subsec:linear}

We start by considering linear regression with Gaussian errors, reproducing results in earlier work \citep[][$\S$3]{SlawskiDiaoBenDavid2019}. In this case, we have 
\begin{equation*}
-\log \phi_i(y_i, \M{x}_i; \bm{\theta})   = -\log \phi(y_i | \M{x}_i; \bm{\theta}) = \frac{1}{2} \log(\sigma^2) +  \frac{1}{2\sigma^2} (y_i - \M{x}_i^{\T} \bm{\beta}), \quad 1 \leq i \leq n,
\end{equation*}
with $\bm{\theta} = (\bm{\beta}, \sigma^2)$. As a result, the M-step for $\bm{\theta}$ based on \eqref{eq:Mstep} reduces to the 
following:
\begin{equation}\label{eq:Mstep_Gaussian}
\wh{\bm{\beta}}^{(t +1)} \leftarrow \argmin_{\bm{\beta}} \sum_{i = 1}^n \{ (1 - \wh{m}_i^{(t)}) (y_i - \M{x}_i^{\T} \bm{\beta})^2 \}, \qquad  \wh{\sigma}^{2\,(t+1)} \leftarrow  \frac{\su (1 - \wh{m}_i^{(t)}) (y_i - \M{x}_i^{\T} \wh{\bm{\beta}}^{(t+1)})}{\sum_{i = 1}^n (1 - \wh{m}_i^{(t)})}.  
\end{equation}
We note that here and below, unless stated otherwise, the intercept is included in the $\{ \M{x}_i \}_{i = 1}^n$.

\subsection{Generalized Linear Regression}
An extension to the class of generalized linear regression models (GLMs, cite) is obtained via the specification  
\begin{equation*}
-\log \phi_{i}(\M{x}_i, y_i; \bm{\theta}) = -\log \phi(y_i | \M{x}_i; \bm{\theta}) = \frac{\psi(\vartheta(\M{x}_i^{\T} \bm{\beta})) - y_i \vartheta(\M{x}_i^{\T} \bm{\beta}) }{\sigma} + c(y_i, \sigma), \quad 1 \leq i \leq n,
\end{equation*}
for a link function $\vartheta$, cumulant $\psi$, scale parameter $\sigma$, and partition function $c$. It is customary to use 
the canonical link in which case $\vartheta$ equals the identity map. Popular examples include {\bfseries (i)} logistic regression with 
$\psi(\cdot) = \log(1 + \exp(\cdot))$ and $\sigma = 1$, and {\bfseries (ii)} Poisson regression with $\psi(\cdot) = \exp(\cdot)$ and 
$\sigma = 1$. A popular example with non-canonical link is  {\bfseries (iii)} Gamma regression with log-link with 
$\vartheta = -\exp(-\cdot)$, $\psi(\cdot) = -\log(-\cdot)$, and $c(y, \sigma) = \frac{\sigma - 1}{\sigma} \log(y) + \frac{\log(\sigma)}{\sigma} + \log(\Gamma(1/\sigma))$. 

In all three cases, the M-step for $\bm{\theta}$ based on \eqref{eq:log_complete} is performed by first obtaining $\wh{\bm{\beta}}^{(t+1)}$ via a (regular) GLM fit with data $\{ (\M{x}_i, y_i) \}_{i = 1}^n$ and 
observation weights $\{ 1 - \wh{m}_i^{(t)}\}_{i = 1}^n$, and then (if necessary) updating the scale parameter $\sigma$
by minimizing the M-step objective \eqref{eq:log_complete} over $\sigma$ with $\bm{\beta}$ fixed to $\bm{\beta}^{(t+1)}$. The latter is a \emph{one-dimensional} optimization problem and hence easy to solve via appropriate routines; in the linear regression case with Gaussian errors, this M-step update reduces to \eqref{eq:Mstep_Gaussian}.


\subsection{Cox PH Regression}
For the (semiparametric) Cox proportional hazards (PH) model, the response variable is given by a right-censored survival time. Accordingly, the data
set is the of the form $\{ ((y_i, \delta_i), \M{x}_i) \}_{i = 1}^n$, where $\delta_i = 1$ if $y_i$ is observed without right-censoring and $\delta_i = 0$ otherwise, $1 \leq i \leq n$. The Cox PH model postulates that 
\begin{equation*}
-\log \phi(y_i|\M{x}_i, \delta_i; \bm{\theta}) = -\delta_i \, \log \lambda(y_i|\M{x}_i;\bm{\theta}) + \Lambda(y_i|\M{x}_i;\bm{\theta}), \quad 1 \leq i \leq n, 
\end{equation*}
with $\lambda(y_i | \M{x}_i;\bm{\theta}) = \lambda_0(y_i) \cdot \exp(\M{x}_i^{\T} \bm{\beta})$ and $\Lambda(y_i | \M{x}_i;\bm{\theta}) = \exp(\M{x}_i^{\T} \bm{\beta}) \Lambda_0(y_i)$, $1 \leq i \leq n$, where $\lambda$ and $\Lambda$ denote the (conditional) hazard and cumulative
hazard functions, respectively, depending on baseline hazard and cumulative hazard functions $\lambda_0$ and $\Lambda_0$, respectively. Here,
$\bm{\theta} = (\bm{\beta}, \lambda_0)$ contains the infinite-dimensional nuisance parameter $\lambda_0$. The M-step for
$\bm{\theta}$ based on is given by 
\begin{equation*}
\min_{\bm{\beta}, \, \lambda_0} \, \left\{ -\sum_{i = 1}^n (1 - \wh{m}_i^{(t)}) \big\{ \delta_i \big[ \log(\lambda_0(y_i)) + \M{x}_i^{\T} \bm{\beta} \big]  + \exp(\M{x}_i^{\T} \bm{\beta}) \Lambda_0(y_i) \big\} \right\}, 
\end{equation*}
where the term inside the curly brackets equal the (full) negative log-likelihood of the Cox model with observation
weights $\{ (1 - \wh{m}_i^{(t)}) \}_{i = 1}^n$. It is well-known that given distinct surival times $\{ y_i \}_{i = 1}^n$ the corresponding profile negative log-likelihood for $\bm{\beta}$ is given by the partial negative log-likelihood 
\begin{equation*}
-\su (1-\wh{m}_i^{(t)}) \log\left \{  \left (\frac{\exp(\M{x}_i^{\T} \bm{\beta})}{\sum_{j \in \mc{R}(y_{i})} \exp(\M{x}_j^{\T} \bm{\beta})} \right )^{\delta_{i}} \right \},
\end{equation*}
where $\mc{R}(y_i) =\{j: y_j \geq y_i \}$, $1 \leq i \leq n$. A minimizer of the above expression can be obtained from any routine for fitting the Cox PH model subject to observation weights, such as the function \texttt{coxph} in the R package \texttt{survival} \citep{survival-package}. Given a minimizer $\wh{\bm{\beta}}^{(t+1)}$, the resulting estimator for the baseline cumulative hazard $\wh{\Lambda}_0^{(t+1)}$ is given by the weighted Breslow estimator 
\begin{equation*}
\wh{\Lambda}_0^{(t+1)}(y)  =  \sum_{i = 1}^{n} \frac{(1 - \wh{m}_i^{(t)}) \mathbb{I}(y_i \leq y)\delta_i}{\sum_{j \in \mc{R}(y_{i})} (1-\wh{m}_j^{(t)})\exp (\M{x}_{j}^{\T}\wh{\bm\beta}^{(t+1)})} \,,
\end{equation*}
and $\wh{\lambda}_0^{(t+1)}$ is obtained as the corresponding piecewise constant function \citep[cf.][]{Breslow1972}. 



\subsection{Unsupervised problems}\label{subsec:up}
To illustrate the unsupervised setting, we consider i) estimation of the covariance matrix of a multivariate normal 
random vector $(\M{x}^{\T} \,\; \M{y}^{\T})^{\T}$, and ii) parameter estimation for a two-way contingency table for categorical variables. Here, the term ``unsupervised" refers to the fact that the roles of $\M{x}$
and $\M{y}$ are symmetric in the sense that there is no distinction between predictor and response variable. 

\noindent {\bfseries i) Multivariate normal data}. The parameter is given $\bm{\theta} = \bm{\Sigma}$, structured according to blocks $\bm{\Sigma}_{\M{x}\M{x}}$, $\bm{\Sigma}_{\M{x}\M{y}}$, 
$\bm{\Sigma}_{\M{y}\M{y}}$ (and $\bm{\Sigma}_{\M{y}\M{x}} = \bm{\Sigma}_{\M{x}\M{y}}^{\T}$), with $\bm{\Sigma}_{\M{x}\M{x}} = \cov(\M{x})$, $\bm{\Sigma}_{\M{y}\M{y}} = \cov(\M{y})$, and $\bm{\Sigma}_{\M{x}\M{y}} = \cov(\M{x}, \M{y})$. For simplicity, 
we assume that $\E[\M{x}]$ and $\E[\M{y}]$ are both zero; in fact, estimation of these quantities is not affected by mismatch
error in the linked file $\{(\M{x}_i, \M{y}_i) \}_{i = 1}^n$, nor is the estimation of the (marginal) covariances 
$\bm{\Sigma}_{\M{x}\M{x}}$ and $\bm{\Sigma}_{\M{y}\M{y}}$. We here slightly depart from the principle according to which
the marginals $f_{\M{x}}$ and  $f_{\M{y}}$ are considered fixed (known or substituted by a plug-in estimator), and instead 
jointly estimate $\M{\Sigma}_{\M{x}\M{x}}$, $\M{\Sigma}_{\M{y}\M{y}}$, and $\M{\Sigma}_{\M{x}\M{y}}$ in a way that is computationally most convenient. Specifically, noting that 
\begin{equation*}
f(\M{x}_i;\bm{\theta}) \times f(\M{y}_i;\bm{\theta}) \propto |\bm{\Gamma}|^{-1/2} \exp\bigg(-\frac{1}{2}\bigg(\begin{array}{c}
     \M{x}_i \\[-1ex]
     \M{y}_i 
\end{array} \bigg)^{\T} \bm{\Gamma}^{-1} \bigg(\begin{array}{c}
     \M{x}_i \\[-1ex]
     \M{y}_i 
\end{array} \bigg) \bigg), \;\, \bm{\Gamma} \coloneq \begin{pmatrix}
\bm{\Sigma}_{\M{x}\M{x}} & \M{0} \\[1ex]
\M{0} &  \bm{\Sigma}_{\M{y}\M{y}}
\end{pmatrix}, \;\, 1 \leq i \leq n,
\end{equation*}
we can consider a modification of the objective in the M-step \eqref{eq:Mstep} by not dropping the terms 
depending on the marginals $f_{\M{x}}$ and $f_{\M{y}}$. The resulting modified expected complete data 
negative (pseudo)-log likelihood then takes the form 
\begin{align}
&-\su \left\{ (1 - \wh{m}_i^{(t)}) \log(\phi_i(\M{x}_i, \M{y}_i;\bm{\theta})) + \wh{m}_i^{(t)} \cdot  f(\M{x}_i;\bm{\theta}) \times f(\M{y}_i;\bm{\theta}) \right\} \notag \\
&\propto  \left\{ -\log|\bm{\Omega}| \bigg(\sum_{i = 1}^n  (1 - \wh{m}_i^{(t)})  \bigg)+ \tr(\bm{\Omega} \M{S}^{(t)})   -\log|\bm{\Psi}|  \bigg(\sum_{i = 1}^n  \wh{m}_i^{(t)} \bigg)  + \tr(\bm{\Psi} \M{S}_{\text{ind}}^{(t)})\right\}, \label{eq:objective_precision}\\
&\bm{\Omega} = \bm{\Sigma}^{-1}, \;\, \bm{\Psi} = \bm{\Gamma}^{-1}, \;\, \M{S}^{(t)} = \sum_{i = 1}^n (1 - \wh{m}_i^{(t)}) \bigg(\begin{array}{c}
     \M{x}_i \\[-1ex]
     \M{y}_i 
\end{array} \bigg) \bigg(\begin{array}{c}
     \M{x}_i \\[-1ex]
     \M{y}_i 
\end{array} \bigg)^{\T}, \;\, \M{S}_{\text{ind}}^{(t)} = \su \wh{m}_i^{(t)}  \begin{pmatrix}
\M{x}_i \M{x}_i^{\T} & \M{0} \\
\M{0} & \M{y}_i \M{y}_i^{\T}
\end{pmatrix} \notag. 
\end{align}
Minimization with respect to $\bm{\Omega}$ and $\bm{\Psi}$ yields the following closed-form updates: 
\begin{equation*}
\bm{\Omega}^{(t+1)} = \left( \M{S}^{(t)} \bigg /  \bigg(\sum_{i = 1}^n  (1 - \wh{m}_i^{(t)})  \bigg)  \right)^{-1}, \quad 
\bm{\Psi}^{(t+1)} = \left( \M{S}_{\text{ind}}^{(t)} \bigg /  \bigg(\sum_{i = 1}^n  \wh{m}_i^{(t)} \bigg) \right)^{-1}. 
\end{equation*}
While these updates do not incorporate the constraint $\bm{\Sigma}_{\M{x}\M{x}} = \bm{\Gamma}_{\M{x}\M{x}}$ and
$\bm{\Sigma}_{\M{y}\M{y}} = \bm{\Gamma}_{\M{y}\M{y}}$, they are straightforward to obtain unlike the situation in which these constraints were imposed explicitly\footnote{These constraints are non-convex since they are formulated in terms of covariance matrices, whereas the objective \eqref{eq:objective_precision} is convex only in the inverse covariance matrix.}. 
\vskip1.5ex
\noindent {\bfseries ii) Two-way contingency tables}. Consider two categorical random variables $\M{x}$ and $\M{y}$
taking values in categories numbered $\{1,\ldots,K \}$ and $\{1,\ldots,L \}$, respectively. Let 
$\theta_{kl} = \p(\M{x} = k, \M{y} = l)$, $1 \leq k \leq K$, $1 \leq l \leq L$, denote the corresponding joint probabilities, and accordingly let $\bm{\theta} = (\theta_{kl})_{k,l}$. Given a linked file $\{ (x_i,  y_i) \}_{i = 1}^n$ whose correctly matched pairs are distributed as $(\M{x}, \M{y})$, we note that the independence assumption in \eqref{eq:mixture_model} implies 
that for mismatched pairs the resulting contribution to the likelihood is given by $f(x_i;\bm{\theta}) \times f(y_i;\bm{\theta}) = \theta_{x_i \, +} \cdot \theta_{+\,y_i} $, where the subscript $+$ indicates summation over the corresponding index. As a notable 
difference from models discussed above, we note that the parameter $\bm{\gamma}$ of the model $h(\cdot\,;\bm{\gamma})$ for the mismatch indicators can no longer be inferred from the data. In fact, consider the case in which $h(\cdot\,;\gamma) = 1 - \gamma$, $\gamma \in (0,1)$, is a constant: it is easy to see that the resulting pseudo-likelihood \eqref{eq:PL} is always maximized by
setting $\gamma = 0$ since the parameters $(\theta_{kl})$ correspond to a saturated model achieving perfect fit regardless 
of the specific $\{ (x_i, y_i) \}_{i = 1}^n$. This issue can be addressed by fixing $\gamma$. Similar to the approach taken
for the multivariate Gaussian model in \eqref{eq:objective_precision}, we propose to work with two separate sets of parameters representing a saturated and an independence model, respectively, and to drop the associated (linear) constraints that would couple these two sets of parameters. Specifically, the expected complete data negative (pseudo)-log likelihood takes the form
\begin{align*}
&-\su \left\{ (1 - \wh{m}_i^{(t)}) \log(\phi_i(x_i, y_i;\bm{\theta})) + \wh{m}_i^{(t)} \cdot  f(x_i;\bm{\theta}) \times f(y_i;\bm{\theta}) \right\} \notag \\
&= -\sum_{k,l} (1 - \wh{m}_{kl}^{(t)}) \log(\pi_{kl}) + \sum_{k,l} \wh{m}_{kl}^{(t)} \log(\psi_{k+} \cdot \psi_{+l}),
\end{align*}
where we note that the $\{ \wh{m}_i \}_{i = 1}^n$ are constant across observations falling into the same cell $(k,l)$
of the associated contigency table, $1 \leq k \leq K$, $1 \leq l \leq L$. In the above display, $\psi_{k+} = \sum_{l} \pi_{kl}$, $1 \leq k \leq K$, and $\psi_{+l} = \sum_{k} \pi_{kl}$, $1 \leq l \leq L$, but for computational simplicity this constraint is dropped when performing the minimization with respect to $\{ \pi_{kl}\}$, $\{ \psi_{k+} \}$, and $\{\psi_{+l} \}$. This minimization amounts to fitting separate saturated and independence models to re-weighted samples with (equivalent) sample
sizes of $\sum_{k,l} (1 - \wh{m}_{kl}^{(t)})$ and $\sum_{k,l} \wh{m}_{kl}^{(t)}$, respectively. Implementation-wise this can be achieved via weighted Poisson regressions, in light of well-known connections between loglinear models for contigency tables 
and Poisson regression \cite[cf., e.g.,][]{Agresti2012}.

\subsection{Modeling the latent mismatch indicator}\label{subsec:mismatch_modeling}
In the preceding sections, we have elaborated on the specifics of various models concerning the relationship between 
$\M{x}$ and $\M{y}$. The second major aspect of modeling concerns the latent mismatch indicators. Since these are binary, the use of a logistic regression model can be considered the standard choice, i.e., in the context of \eqref{eq:mixture_likelihood} and \eqref{eq:mixture_likelihood_regression} 
\begin{equation*}
\p(m_i = 0|\M{z}_i) = h(\M{z}_i;\bm{\gamma}) = \frac{\exp(\M{z}_i^{\T} \bm{\gamma})}{1 + \exp(\M{z}_i^{\T} \bm{\gamma})}, \quad 1 \leq i \leq n. 
\end{equation*}
If no auxiliary covariates 
$\{ \M{z}_i \}_{i=1}^n$ informative of the match status are available, an intercept-only model can be employed which is equivalent to assuming a constant mismatch rate (regardless of the choice of the link function). It is worth stressing that despite the similarities in modeling, estimation of the parameters is more challenging than in (plain) binary regression since the $\{ m_i \}_{i = 1}^n$ are not observed. To faciliate parameter estimation, it can be helpful to integrate prior knowledge about the underlying mismatch rate by imposing a linear constraint on the average of the linear predictor of the form  $(\frac{1}{n} \su \M{z}_i)^{\T} (-\bm{\gamma}) \leq b$, where $b \in \R$ corresponds to the logit of the assumed mismatch rate. Such a constraint can be incorporated in a straightforward manner within the approach to inference presented in $\S$\ref{sec:inference} above.  


\subsection{Other applications}
The proposed approach generalizes in a straightforward fashion to various other settings involving mismatch error that have been considered in recent literature, including spherical regression \citep{Shi2021} and multivariate linear regression \citep{SlawskiBenDavidLi2019}. We omit detailed discussions for the sake of brevity.  Extensions to linear mixed effect models and M-estimation are possible though more intricate and are left for future work \citep{FabriziSalvatiSlawski2023}.

\section{Miscellaneous details and extensions}\label{sec:misc}
This section serves as an addendum to the two preceding sections $\S$\ref{sec:inference} and $\S$\ref{sec:examples}, filling in additional details on the estimation of marginal PDFs and outlining two specific extensions of the basic framework.  

\subsection{Estimation of marginal PDFs}\label{subsec:estimation_marginals}
In section $\S$\ref{sec:inference}, the marginal densities $f_{\M{x}}$ and $f_{\M{y}}$ were treated as known quantities. In practice, this is not the case even though estimation is considered less of a challenge given that mismatch error affects estimation 
(of parameters) of the joint distribution but not of the marginals. In general, we distinguish between two approaches: i) {\em plug-in estimation} and ii) {\em integrated estimation}. In the former approach, the marginal PDFs are estimated beforehand and plugged in as replacement for the corresponding population quantities; in the second approach, the marginal PDFs are updated along with the parameter $\bm{\theta}$ of primary interest.    

i) The plug-in approach reduces to plain density estimation of $f_{\M{y}}$ (and also of $f_{\M{x}}$ outside regression setups),  and various methods ranging 
from fully non-parametric to parametric are available to perform this task. Particular examples include kernel density estimation or the use of empirical probability mass functions if the range of the associated random variable is discrete and small in size. Note that while in the plug-in approach the marginal PDFs are not updated during the EM iterations, they enter in the E-step \eqref{eq:Estep_general} as well as in the evaluation of the pseudo-likelihood at the iterates $\{ (\wh{\bm{\theta}}^{(t)}, \, \wh{\bm{\gamma}}^{(t)}) \}_{t \geq 1}$.   

ii) In the integrated approach, $f_{\M{x}}$ and $f_{\M{y}}$ are updated with $\bm{\theta}$. If correctly paired observations are \emph{i.i.d.}~with joint PDF $f_{\M{x},\M{y}}(\cdot,\cdot;\bm{\theta})$, the relationships $f_{\M{x}}(\cdot;\bm{\theta}) = \int f_{\M{x},\M{y}}(\cdot,\M{y};\bm{\theta}) \, d\M{y}$ and $f_{\M{y}}(\cdot) = \int f_{\M{x},\M{y}}(\M{x},\cdot;\bm{\theta}) \, d\M{x}$ prompt updates along with $\bm{\theta}$. Such updates typically arise in unsupervised settings (cf.~$\S$\ref{subsec:up}). By contrast, in standard fixed design regression setups, $f_{\M{y}}$ can
be expressed as the finite mixture



%
\begin{equation*}
f_{\M{y}}(\cdot;\bm{\theta}) = \int \phi(\cdot|\M{x};\bm{\theta}) \; dP(\M{x}) = \frac{1}{n} \su \phi(\cdot|\M{x}_i;\bm{\theta}),
\end{equation*}
where $P$ denotes the atomic measure with atoms $\{ \M{x}_i \}_{i = 1}^n$ each having mass $1/n$. For example, in classical linear regression with i.i.d.~Gaussian errors (cf.~$\S$\ref{subsec:linear}), the above mixture density becomes the Gaussian location mixture 
\begin{equation}\label{eq:mixture_y}
f_{\M{y}}(\cdot;\bm{\beta},\sigma) = \frac{1}{n} \su \frac{1}{\sigma} \varphi \left(\frac{\cdot - \M{x}_i^{\T} \bm{\beta}}{\sigma} \right),
\end{equation}
where $\varphi$ denotes the PDF of the $N(0,1)$-distribution. Incorporating this into the EM approach in \ref{subsec:optimization} would break the simplicity of the updates, which appears too much of a price to pay
given only minor gains in statistical efficiency over the plug-in approach in which $f_{\M{y}}$ could simply be replaced by 
a kernel density estimator based on the $\{ y_{i} \}_{i= 1}^n$\cite[see][for a related discussion concerning Gaussian design linear regression]{SlawskiDiaoBenDavid2019}. As a compromise, an initial kernel density estimator can be replaced by the representation in \eqref{eq:mixture_y} with $(\bm{\beta}, \sigma)$ substituted by 
estimates $(\wh{\bm{\beta}}, \wh{\sigma})$ obtained from a first round of EM iterations. 

A simplified approach for GLMs is to model $f_{\M{y}}$ in terms of an intercept-only GLM (and potentially a scale parameter). In particular, this is relevant to binary GLMs in which case the intercept is simply a one-to-one transformation of 
$\p(y_i = 1)$, $1 \leq i \leq n$. In Normal GLMs, if the predictor variables follow a Normal distribution, then $f_{\M{y}}$  is also a Normal distribution with unknown mean (intercept) and standard deviation (scale parameter). It is justifiable to adopt the latter model at least as a simple approximation outside the setting of Normal predictors \citep[cf.][]{SlawskiDiaoBenDavid2019}.
\vskip1ex
\noindent {\em Estimation of $f_{\M{y}}$ for the Cox PH model}. The likelihood contribution for
observation $i$ pertaining to the marginal distribution of the outcome variable is given by 
$\exp(-\Lambda_{\M{y}}(y_i)) \lambda_{\M{y}}(y_i)^{\delta_i}$ (recall that $\delta_i = 1$ if observation $i$ is not right-censored), $1 \leq i \leq n$, where $\lambda_{\M{y}}$ and $\Lambda_{\M{y}}$ denote the hazard and cumulative hazard function
associated with $f_{\M{y}}$. We propose to estimate the cumulative hazard function via the Nelson-Aalen estimator, and take the associated jump heights of the resulting step function as a (piecewise constant) estimator of the hazard function.  

\subsection{Test for zero mismatch error}
The null hypothesis corresponding to such a test can be formulated as 
$\text{H}_0:\; h \equiv 1$. The corresponding test cannot be performed based 
on standard techniques (such as a likelihood ratio test) since the model 
associated with the null in general constitutes an element of the boundary 
of the parameter space, and hence one of the regularity assumptions that underlies these techniques is not satisfied. This issue is well-known in the context of a plain mixture models \citep[e.g.][]{Chen2009} when the goal is to conduct statistical tests to determine an appropriate number of mixture components. Recently, \cite{Wasserman2020} have proposed a test based on sample splitting that can be employed in such ``non-regular" settings. The test statistic proposed therein (the so-called split likelihood ratio statistic) is (almost) universally applicable and ensures finite-sample control of the type I error. We here outline how this approach can be applied in conjunction with the model under consideration in the present paper. 

\noindent {\em Step I}. Divide the entire data set into two disjoint subsets $\mc{D}_0$ and $\mc{D}_1$ (of roughly equal size). 
\vskip1ex
\noindent {\em Step II}. Denote by $\wt{\mc{L}}_0(\bm{\theta})$ the (pseudo)-likelihood associated with
$\mc{D}_0$ {\em after fixing $h \equiv 1$}, and compute $\wh{\bm{\theta}}_0 = \argmax_{\bm{\theta}} \mc{L}_0(\bm{\theta})$.   
\vskip1ex
\noindent {\em Step III}. Denote by $\mc{L}_1(\bm{\theta}, \bm{\gamma})$ the (pseudo)-likelihood associated with
$\mc{D}_1$, and compute $(\wh{\bm{\theta}}_1,  \wh{\bm{\gamma}}_1) = \argmax_{(\bm{\theta}, \bm{\gamma})} \mc{L}_1(\bm{\theta}, \bm{\gamma})$. 
\vskip1ex
\noindent {\em Step IV}. Compute $T = \log \mc{L}_{0}(\wh{\bm{\theta}}_1, \wh{\bm{\gamma}}_1) - \log \wt{\mc{L}}_{0}(\wh{\bm{\theta}}_0)$, where $\mc{L}_0(\bm{\theta}, \bm{\gamma})$ the (pseudo)-likelihood associated with
$\mc{D}_0$ ({\em without constraining} $h$). Reject the null hypothesis at significance level $\alpha \in (0,1)$ if 
$T > \log(1/\alpha)$. Accordingly, a p-value can be defined as $e^{-T}$. 
\vskip2ex
One particular criticism of this approach is that its results depends on the specific subsets $\mc{D}_0$ and $\mc{D}_1$. Remediations such as swapping the roles of $\mc{D}_0$ and $\mc{D}_1$ and re-computing the above statistic for multiple
splits are discussed in \cite{Wasserman2020}. 



\subsection{Bayesian Inference}\label{subsec:Bayesian}
There are situations where it can be useful to recast the proposed approach in a Bayesian framework to facilitate
inference. In particular, this is the case in which regularization is imperative to deal with a large number of 
parameters. For example, RL is often performed after blocking, and each block may be associated with its own mismatch rate. If the number of blocks is large, placing a prior on the block-wise mismatch rates (e.g., a Beta prior) is appropriate for ``borrowing strength" across blocks when estimating these rates. We do not pursue this case further, and instead consider
another scenario of interest, namely smooth curve fitting via penalized splines. Specifically, the ``roughness penalty" \citep{Green1993} is realized via an (improper) Gaussian prior on the spline coefficients. This connection facilitates the data-driven choice of the level of smoothing, which is particularly helpful when other criteria such as the GCV \citep{Craven1978} are not easily applicable. Moreover, subsequent inference (e.g., pointwise standard errors for the regression curve) becomes rather straightforward within a Bayesian framework.

Specifically, we consider the following setup expressed in a hierarchical Bayes formulation: 
\begin{alignat}{2}\label{eq:splinemodel}
&f(\alpha) \propto 1, \qquad f(\sigma^2) \propto (\sigma^2)^{-1}, \quad \;\;\,&&f(\tau^2) \propto (\tau^2)^{-1} \notag\\
&\{ m_i \}_{i = 1}^n|\alpha \overset{\text{i.i.d}}{\sim} \text{Bernoulli}(\alpha), \quad \;\; \,&&f(\bm\beta | \tau^2) \propto  (\tau^2)^{-r/2} (\det \M{S}^+)^{1/2} \exp \left(-\frac{1}{2\tau^2}\bm\beta^{\T} \M{S} \bm\beta \right) \\
& y_i | x_i, \{m_i = 0 \}, \bm\beta, \sigma^2 \sim N(s_{\bm\beta}(x_i), \sigma^2), \quad \;\;\, && y_i|\{m_i = 1\} \sim f_y, \qquad i=1,\ldots,n, \notag
\end{alignat}
where for $\bm\beta \in \R^d$, the function $s_{\bm\beta}(x) = \sum_{j = 1}^d \beta_j B_j(x)$ is a cubic spline expansion with
coefficients $\bm\beta = (\beta_j)_{j = 1}^d$ and basis functions $\{ B_j \}_{j = 1}^d$ on some interval $[a, b]$ covering
the range of the predictor variable. To keep the setup simple, the mismatch indicators are assumed to be i.i.d.~Bernoulli random variables, but an additional layer can be added to model these indicators conditional on covariates. The prior 
$f(\bm\beta | \tau^2)$ is an established construct in the spline literature, \citep[cf., e.g.,][]{Ruppert2003}; in \eqref{eq:splinemodel}, $^+$ denotes the Moore-Penrose pseudo-inverse, and $r$ equals the rank of the roughness penalty matrix $\M{S}$. Note that improper Gamma priors are placed on $\sigma^2$ and $\tau^2$, with $\sigma^2 / \tau^2$ corresponding to the effective smoothing parameter. 
\begin{figure}
\hspace*{-.5ex}\includegraphics[height = 0.4\textwidth]{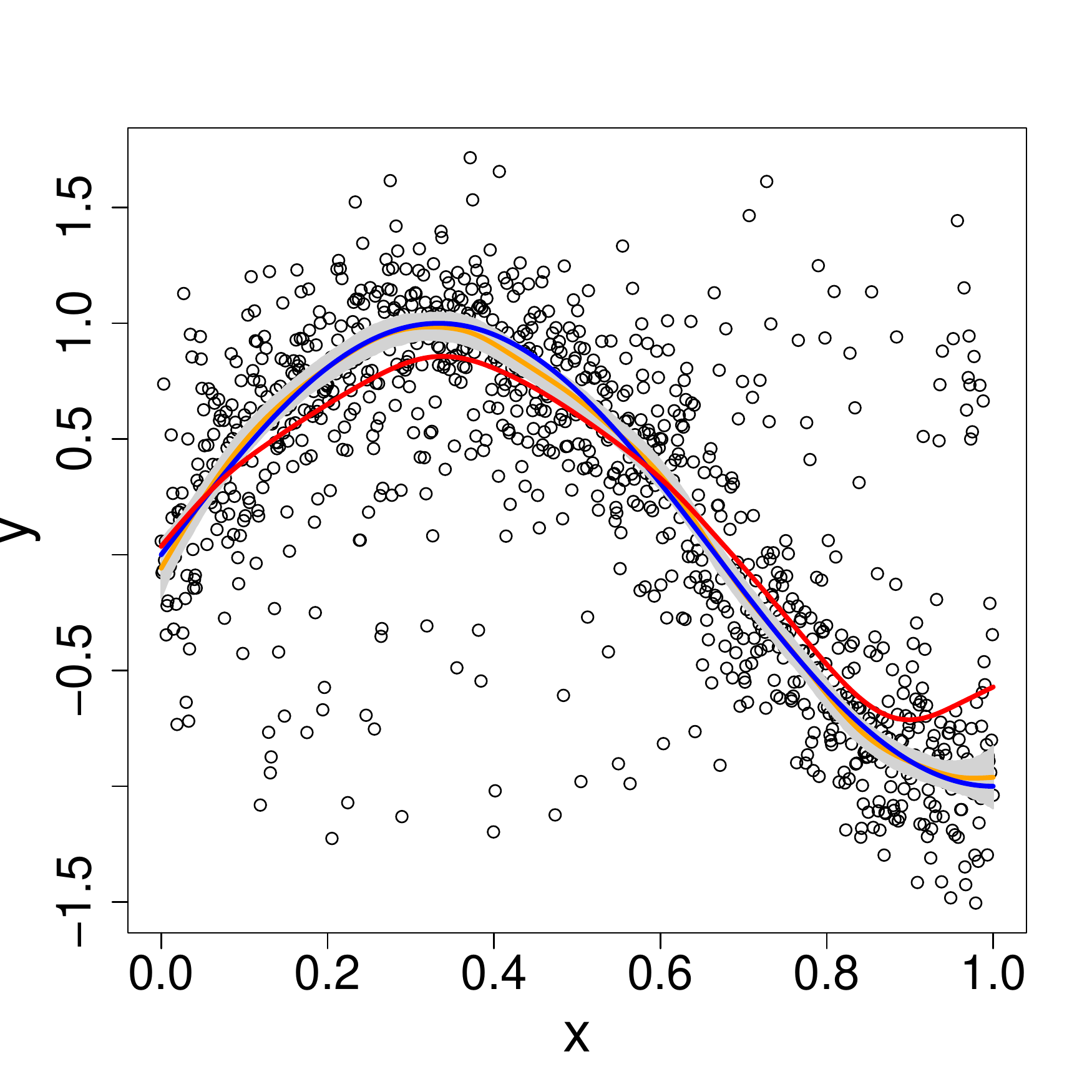}
\hspace*{-1ex}\begin{minipage}{.55\textwidth}
\vspace*{-36ex}{\footnotesize \begin{tabular}{|l|l|l|l|l|}
\hline
         & naive & oracle & posterior mean & 95\% credible interval  \\
         \hline
$\alpha$    & -- &  --       &  .1966 (0.2$^{\dagger}$) & [.155, .235]\\
$\sigma^2$    & .222 & .061     &  .0630  ($0.0625^{\dagger}$) & [.055, .072]\\
$\sigma^2 / \tau^2$ & 1088.6$^*$ & 413.1$^*$ & 383.9 & [89.4, 911.5]\\
\hline
\end{tabular}}
\vskip1ex
{\footnotesize $^*$: smoothing parameter selected by \texttt{mgcv}. \\
$\dagger$: true parameter values.} 
\end{minipage}
\vspace*{-3ex}
\caption{Left: Realizations from a noisy sine function (blue) with 20\% random mismatches. Red: unadjusted spline fit; orange (hardly distinguishable from blue) adjusted spline fit (point-wise posterior mean). Grey-shaded region: 95\% point-wise credible bands.}\label{fig:spline_simulation}
\end{figure}

A particularly convenient feature of the model specification \eqref{eq:splinemodel} is that posterior inference can be performed via Gibbs sampling with standard distributions for the full conditionals. Specifically, we have 
{\small \begin{align*}
&\alpha|\{m_i \}_{i = 1}^n \sim \text{Beta}\left(\sum_{i = 1}^n m_i + 1, n - \sum_{i = 1}^n m_i + 1 \right), \quad m_i|\{y_i, x_i \}, \alpha, \bm\beta, \sigma^2  \sim \text{Bernoulli}(\pi_i), \; \; 1 \leq i \leq n,\\
&\sigma^2 | \{y_i, x_i, m_i \}, \bm\beta  \sim \text{Inverse-Gamma}\left( \left(n - \sum_{i = 1}^n m_i \right) \big/ 2, \sum_{i = 1}^n (y_ i - s_{\bm\beta}(x_i))^2 \big / 2 \right) \\
&\tau^2 | \beta \sim \text{Inverse-Gamma}(r/2, \bm\beta^{\T} \M{S} \bm\beta / 2), \\
&\bm\beta | \{y_i, x_i, m_i\}, \sigma^2, \tau^2 \sim N(\wt{\bm\beta}, \wt{\bm\Sigma}), \qquad \wt{\bm\beta} \coloneq \left( \sum_{i = 1}^n (1 - m_i) \M{b}_i \M{b}_i^{\T}  + \frac{\sigma^2}{\tau^2} \M{S} \right)^{-1} \sum_{i= 1}^n (1 - m_i) \M{b}_i y_i, 
\\
&\qquad \qquad \qquad \qquad \qquad \qquad \qquad \quad \;\;\wt{\bm{\Sigma}} \coloneq \sigma^2 \left(\sum_{i = 1}^n (1 - m_i) \M{b}_i \M{b}_i^{\T} + \frac{\sigma^2}{\tau^2} \M{S} \right)^{-1}, 
\end{align*}}
where $\pi_i = \alpha f_y(y_i) / \{ \alpha f_y(y_i) + (1 - \alpha) f_{y|x}(y_i) \}$ with
$f_{y|x}(y_i)$ representing the PDF of $y_i | x_i, \{ m_i = 0 \}$ evaluated at $y_i$, $1 \leq i \leq n$.

As an illustration, we simulate data $y_i = \sin(\frac{3}{2} \pi x_i ) + 0.25 \cdot \eps_i$, $\{\eps_i \}_{i =1}^n \overset{\text{i.i.d.}}{\sim} N(0,1)$, $x_i = (i-1)/(n - 1)$, $1 \leq i \leq n = 1,000$, and randomly
shuffle 20\% of the $(x_i, y_i)$-pairs. We then fit a cubic spline (25 equi-spaced knots) to the resulting data, without any adjustment, using the R package \texttt{mgcv} \citep{mgcv-package} as well as with the approach outlined above (a kernel density estimator based on the $\{ y_i \}$ is used for $f_y$). The results
are shown in Figure \ref{fig:spline_simulation}. It can be seen that the proposed approach successfully remediates the effect of mismatches, and that all model parameters are estimated accurately. Moreover, the level of smoothing with the proposed approach aligns closely with the level smoothing for correctly matched data, whereas ignoring mismatches altogether yields an oversmoothed fit. In this regard, the Bayesian perspective is particularly helpful since it is unclear how to select the smoothing parameter in a penalized likelihood framework and conduct subsequent statistical inference.

\section{Simulations}\label{sec:simulations}
We here present the results of a set of simulation studies to investigate the empirical performance 
of our approach in a series of different scenarios, including the case of (partial) model misspecification 
and gentle violations of some of the underlying assumptions listed at the beginning of $\S$\ref{sec:meth}. 

\subsection{Poisson GLM}
In the first set of simulations, we consider a single predictor variable $x$ and an outcome variable $y$
following a Poisson distribution with $\E[y|x] = \exp(\beta_0^* + \beta_1^* x)$. Specifically, we consider
a fixed design with $\{ x_i \}_{i = 1}^n$, $n = 1,000$ uniformly spaced between $1$ and $5$ and $\beta_0^* =  0.5, 
\, \beta_1^* = 2$. 
\vskip.5ex
\noindent Several settings are considers for the mismatch indicator:

\noindent {\em Constant}. The $\{ m_i \}_{i = 1}^n$ are sampled i.i.d.~from a Bernoulli distribution with 
probability of success $\alpha^* \in \{0,0.05, \ldots, 0.3\}$. Given the $\{ m_i \}_{i = 1}^n$ the corresponding 
subset of the $\{ y_i \}_{i = 1}^n$ is permuted according to a right circular shift. 
\vskip.3ex
\noindent {\em Blockwise}. The data set is sub-divided into four subsets (``blocks") of equal size. Within each of the blocks
the mismatch rates are constant (equal to $0$, $0.1$, $0.4$, $0.6$, respectively), and within each of the blocks
the procedure described under {\em Constant} is applied. 
\vskip.3ex
\noindent {\em Logistic}. In an attempt to mimic the situation in which output from probabilistic record linkage
is available to the data analyst operating on linked data, we consider auxiliary data 
$z_i = \text{logit}(p_i)$, where the $p_i$'s take the place of match ``probabilities" assigned to the $i$-th linked pair (as potentially supplied by a record linkage procedure), $1 \leq i \leq n$. Here, the $\{ p_i \}_{i = 1}^n$ are drawn i.i.d.~from a Beta distribution with parameters $4.5$ and $0.5$. Subsequently, the $\{ m_i \}_{i = 1}^n$ are generated according to the logistic model 
\begin{equation}\label{eq:logistic}
\text{logit}\{ \p(m_i = 0 | z_i)\} = \gamma_0^* + \gamma_1^* z_i, \quad 1 \leq i \leq n,
\end{equation}
where $\gamma_0^* = -0.5$ and $\gamma_1^* = 1$. Given the $\{ m_i \}_{i = 1}^n$, the $y_i$'s are permuted as described under {\em Constant}.\\
For all three settings, the model for the mismatch indicator is specified accordingly when applying our approach. The marginal density $f_y$ is estimated via a kernel density estimator with rectangular kernel and bandwidth fixed to $100$ throughout all simulations.  
\vskip1ex
\noindent In addition to the above settings, we consider three further settings associated with model 
misspecification and/or violation of assumptions. We conduct 10k replications per setting. 
\vskip.3ex
\noindent {\em Mis-$y$}. The linear predictor in the Poisson model is mis-specified. Instead 
of a linear model, a quadratic model in $x$ is used to generate the $y$'s. Specifically, 
$\E[y|x] = \beta_0^* + \beta_1^* x + \beta_2^* x^2$ with $\beta_2^* = 0.05$. This model for 
$y$ is combined with the constant mismatch rate scenario described above. 
\vskip.3ex
\noindent {\em Mis-$m$}. As a modification of the setting ``logistic" above,  model \eqref{eq:logistic}
is changed as follows: 
\begin{equation}\label{eq:logistic_mis}
\text{logit}\{ \p(m_i = 0 | z_i)\} = 0.5 (\gamma_0^* + \gamma_1^*) z_i \mathbb{I}(2 \leq x_i \leq 4), \quad 1 \leq i \leq n,
\end{equation}
When applying our approach, we instead fit a logistic model linear in $z$ in accordance \eqref{eq:logistic}. Note that in addition to using a mis-specified model for the mismatch indicator, the fact that the mismatch indicator depends on $x$ also constitutes a violation of the independence assumption {\bfseries (A1)}. 
\vskip.3ex
\noindent {\em Mis-ind}. In this setting the $x_i$'s are partitioned into 50 blocks of size 20. Within each block, 
the $x_i$'s are simulated according to a Gaussian copula inducing dependence between each set of 20 $x_i$'s whose marginal distribution is uniform on $[1,5]$. The associated covariance matrix of the Gaussian copula is taken as the equi-correlation matrix with unit diagonal elements and off-diagonal elements equal to $0.5$. A constant mismatch rate is assumed within each block, and the $y_i$'s for which $m_i = 1$ are permuted as described under 
{\em Constant}. Note that this simulation design violates assumption {\bfseries (A2)}, part (IND).

\begin{table}
{\sf Constant}
\vspace*{-3ex}
\begin{center}
\footnotesize
\addtolength{\tabcolsep}{-1.5pt}   
\vspace{-1ex}
\begin{tabular}{|l|l|ccc|ccc|ccc||ccc|ccc|ccc|}
\hline
\multicolumn{1}{|c|}{}& \multicolumn{1}{|c|}{} & \multicolumn{3}{c|}{$\wh\beta_0$}&  \multicolumn{3}{c|}{$\wh\beta_1$}& \multicolumn{3}{c||}{$\wh\gamma$} & \multicolumn{3}{|c|}{$\wh\beta_0$ \textsf{(mis-ind)}} & \multicolumn{3}{c|}{$\wh\beta_1$ \textsf{(mis-ind)}} & \multicolumn{3}{c|}{$\wh\gamma$ \textsf{(mis-ind)}} \\ \hline
$\alpha^*$& & RB & SD & CG &   RB & SD & CG & RB & SD & CG & RB & SD & CG & RB & SD & CG  & RB & SD & CG \\ \hline 
     0    & {\bfseries \textsf{w/}} & $2\textsf{e}^{-5}$ & $4\textsf{e}^{-3}$ &  $.95$ & $8\textsf{e}^{-6}$ &  $1\textsf{e}^{-3}$ & $.95$ &  &  & & $6\textsf{e}^{-5}$ & $4\textsf{e}^{-3}$ & $.95$ & $4\textsf{e}^{-6}$ & $1\textsf{e}^{-3}$ & $.95$ & & & \\
          & {\bfseries \textsf{w/o}}  & $2\textsf{e}^{-5}$ & $2\textsf{e}^{-3}$ & $.95$ & $8\textsf{e}^{-6}$ &  $1\textsf{e}^{-3}$ & $.95$ &  &  &  & $6\textsf{e}^{-5}$ & $4\textsf{e}^{-3}$ & $.95$ &  $4\textsf{e}^{-6}$ & $1\textsf{e}^{-3}$ & $.95$ & & & \\
             \hline
                0.05   & {\bfseries \textsf{w/}}    & $4\textsf{e}^{-5}$  & $4\textsf{e}^{-3}$ & $.95$ & $2\textsf{e}^{-6}$ & $1\textsf{e}^{-3}$ & $.95$ & $-.01$  & $.15$ & $.95$ & $9\textsf{e}^{-5}$ & $4\textsf{e}^{-3}$  & $.95$ & $5\textsf{e}^{-6}$ & $9\textsf{e}^{-4}$ & $.95$ & $-.02$ & $.16$ & $.94$ \\
             & {\bfseries \textsf{w/o}} & $2.4$ & $.34$ & $.06$ & $-.13$ & $.08$& $.07$ & & & & $1.1$ & $.20$ & $.27$ & $-.06$ & $.05$ & $.27$ & & & \\
\hline          
    0.10     & {\bfseries \textsf{w/}}   & $1\textsf{e}^{-5}$  & $5\textsf{e}^{-3}$ & $.95$& $1\textsf{e}^{-6}$ & $1\textsf{e}^{-3}$ & $.95$ & $-.01$ & $.11$ & $.95$ & $1\textsf{e}^{-4}$ & $5\textsf{e}^{-3}$ & $.95$ & $5\textsf{e}^{-6}$ & $1\textsf{e}^{-3}$ & $.95$ & $-.03$  & $.15$ & $.92$  \\
             & {\bfseries \textsf{w/o}}   & $4.2$ & $.38$ & $0$  & $-.23$ & $.09$ & $0$ & & & & $2.1$ & $.30$ & $.03$ & $-.12$ & $.07$ & $.03$ & &   & \\
              \hline          
    0.15   & {\bfseries \textsf{w/}}      & $1\textsf{e}^{-4}$  & $5\textsf{e}^{-3}$  & $.95$ & $4\textsf{e}^{-6}$ & $1\textsf{e}^{-3}$ & $.95$ & $-.01$ & $.09$ & $.95$ & $2\textsf{e}^{-4}$  & $5\textsf{e}^{-3}$ & $.95$ & $1.2$ & $1\textsf{e}^{-3}$ & $.95$ & $-.04$ & $.11$ & $.90$  \\
           & {\bfseries \textsf{w/o}}   & $5.7$ & $.37$ & $0$ & $-.32$ &  $.09$ & $0$ & & & & $3.0$ & $.28$ & $0$ & $-.17$ & $.06$ & $0$ &  &  &  \\
          \hline
    0.20  & {\bfseries \textsf{w/}}    & $6\textsf{e}^{-5}$ & $5\textsf{e}^{-3}$ & $.95$ & $3\textsf{e}^{-6}$ & $1\textsf{e}^{-3}$ & $.95$ & $-.01$ & $.08$ & $.95$ & $2\textsf{e}^{-4}$ & $5\textsf{e}^{-3}$ & $.95$ & $1\textsf{e}^{-5}$ & $1\textsf{e}^{-3}$ & $.95$ & $-.04$ & $.08$  & $.90$ \\
          & {\bfseries \textsf{w/o}}    & $7.0$ & $.35$ & $0$ & $-.40$ &  $.08$ & $0$ & & & & $3.8$ & $.27$ & $0$ & $-.21$ & $.06$ & $0$ &  &  &  \\
                       \hline
    0.25  & {\bfseries \textsf{w/}}     & $3\textsf{e}^{-4}$ & $5\textsf{e}^{-3}$ & $.95$ & $2\textsf{e}^{-5}$ & $1\textsf{e}^{-3}$ & $.95$ &  $-.02$ &  $.07$ & $.95$ & $2\textsf{e}^{-4}$ & $5\textsf{e}^{-3}$ & $.95$ & $1\textsf{e}^{-5}$ & $1\textsf{e}^{-3}$ & $.95$ & $-.05$ & $.08$ & $.88$ \\
           & {\bfseries \textsf{w/o}}  & $8.0$ & $.33$ & $0$ & $-.46$ & $.08$ & $0$ & & & & $4.5$ & $.27$ & $0$ & $-.26$ & $.06$ & $0$ & & & \\
                           \hline
    0.3   & {\bfseries \textsf{w/}}    & $1\textsf{e}^{-4}$ & $5\textsf{e}^{-3}$ &  $.95$ & $8\textsf{e}^{-6}$ & $1\textsf{e}^{-3}$ & $.95$ & $-.02$ & $.07$ & $.95$ & $5\textsf{e}^{-4}$ & $5\textsf{e}^{-3}$ & $.95$ & $3\textsf{e}^{-5}$ & $1\textsf{e}^{-3}$ & $.95$ & $-.07$ & $.07$ & $.87$ \\
          & {\bfseries \textsf{w/o}}    & $8.9$ & $.31$ & $0$ & $-.51$ & $.08$ & $0$ &  & & & $5.2$ & $.27$ & $0$ & $-.30$ & $.06$ & $0$ & & & \\
             \hline
\end{tabular}
\end{center}
\vskip1ex
{\sf Blockwise}
\begin{flushleft}
\footnotesize
\addtolength{\tabcolsep}{-1pt}    
\vspace{-1ex}
\begin{tabular}{|l|ccc|ccc|ccc|ccc|ccc|ccc|}
\hline
\multicolumn{1}{|c|}{} & \multicolumn{3}{c|}{$\wh\beta_0$}&  \multicolumn{3}{c|}{$\wh\beta_1$}& \multicolumn{3}{c|}{$\wh\gamma_1$} & \multicolumn{3}{|c|}{$\wh\gamma_2$} & \multicolumn{3}{c|}{$\wh\gamma_3$ } & \multicolumn{3}{c|}{$\wh\gamma_4$}  \\ \hline
& RB & SD & CG &   RB & SD & CG & RB & SD & CG & RB & SD & CG & RB & SD & CG  & RB & SD & CG \\ \hline 
       {\bfseries \textsf{w/}} & $3\textsf{e}^{-4}$ & $6\textsf{e}^{-3}$  & $.95$ & $2\textsf{e}^{-5}$ & $1\textsf{e}^{-3}$ & $.95$ & -- & $^{\bm{\star}}$ & -- & $.21$ & $.25$ & $.91$ & $.09$ & $.13$ & $.90$ & $.04$ & $.13$ & $.94$ \\
           {\bfseries \textsf{w/o}}  & $2.8$ &  $.12$ & $0$ & $-.16$ & $.03$ & $0$ & & & & & & & & & & & & \\
             \hline
\end{tabular}
\vskip1ex
$^{\bm{\star}}$: not reported because the mismatch rate in block one is zero.  
\end{flushleft}
\vspace*{-2ex}
{\sf Logistic}
\begin{flushleft}
\footnotesize
\vspace{-1ex}
\begin{tabular}{|l|l|ccc|ccc|ccc|ccc|}
\hline
\multicolumn{1}{|c|}{}& \multicolumn{1}{|c|}{} & \multicolumn{3}{c|}{$\wh\beta_0$}&  \multicolumn{3}{c|}{$\wh\beta_1$}& \multicolumn{3}{c|}{$\wh\gamma_0$} & \multicolumn{3}{|c|}{$\wh\gamma_1$} \\ \hline
& & RB & SD & CG &   RB & SD & CG & RB & SD & CG & RB & SD & CG \\ \hline 
         \eqref{eq:logistic} & {\bfseries \textsf{w/}} & $1\textsf{e}^{-4}$ & $5\textsf{e}^{-3}$ & $.95$ & $8\textsf{e}^{-6}$  & $1\textsf{e}^{-3}$ & $.95$ & $.02$ & $.22$ & $.95$ & $5\textsf{e}^{-3}$ & $.10$ & $.95$ \\
          & {\bfseries \textsf{w/o}}  & $4.9$ & $.33$ & $0$  & $-.28$ & $.08$ & $0$ &  & & & & &  \\
             \hline
                \textsf{mis-m} \eqref{eq:logistic_mis}  & {\bfseries \textsf{w/}}    & $3\textsf{e}^{-5}$ & $5\textsf{e}^{-3}$ & $.95$ & $1\textsf{e}^{-6}$ & $1\textsf{e}^{-3}$  & $.95$ & -- & $^{\bm\dagger}$ & -- & -- & $^{\bm\dagger}$ & -- \\
             & {\bfseries \textsf{w/o}} & $.58$ & $.05$ & $0$ & $-.03$ & $.01$ & $0$ & & & & &  & \\
\hline          
\end{tabular}
\vskip1ex
$^{\bm{\dagger}}$: not reported because the corresponding part of the model is misspecified.
\end{flushleft}
\vspace*{-2.5ex}
\caption{Results of the simulation study for the Poisson GLM based on 10k replications. Note that instead of mismatch rates $\alpha^*$, we estimate the parameter $\gamma^* = \log((1- \alpha^*)/\alpha^*)$. RB -- relative bias; SD -- standard deviation; CG -- coverage rate of confidence intervals. {\bfseries \textsf{w/}}: with adjustment (proposed approach); {\bfseries \textsf{w/o}}: without adjustment, i.e., the naive estimator (plain GLM) ignoring mismatch error.}\label{tab_sim_poisson}
\end{table}

{\em Results}. Under correct model specifications, the proposed approach largely performs as expected. Confidence interval coverage levels achieve the nominal 95\% for all model parameters, with slight under-coverage for the parameter $\gamma^*$ (the logit of the correct match rate $1 - \alpha^*$) only under the {\em Blockwise} setting; we suspect that this might be attributable to the reduced sample size in each block. Table \ref{tab_sim_poisson} also shows that the impact of mismatches becomes noticeable once 10\% of the observations are incorrectly matched. Plain GLM estimates for the regression parameters follow a typical pattern of attenuation characterized by an inflated intercept 
and a reduced slope. By contrast, with the proposed adjustment, estimation of the regression parameters is not visibly affected. In addition, substantial losses in statistical efficiency in the absence of mismatches ($\alpha^* = 0$) are not observed either.  

In the presence of model mis-specification and/or violation of assumptions, the regression coefficients are still estimated accurately and confidence level coverage is maintained, with the exception of setting {\em Mis-$y$} in which the linear predictor is misspecified. For the latter setting, performance is evaluated in terms of the Kullback-Leibler (KL) divergence between the $\{ \mu_i^* = \E[y_i | x_i] \}_{i = 1}^n$ and the corresponding estimates $\{ \wh{\mu}_i \}_{i = 1}^n$ given an incorrectly specified linear predictor; the KL divergence with adjustment ranges between $6.0$ ($\alpha^* = 0$) and $7.8$ ($\alpha^* = 0.3$) after adjustment, whereas without adjustment the KL divergence equals 
only $1.9$ for $\alpha^* = 0$ but then jumps to $208$ for $\alpha^* = 0.05$ and increases to almost $3.3k$ for $\alpha^* = 0.3$. While it is found that the different forms of mis-specifications studied here do not have a noticeable impact concerning estimation of the $x$-$y$ relationship, estimation of the model parameters pertaining to the latent mismatch indicators $\{ m_i \}_{i = 1}^n$ is affected more noticeably. For instance, in the setting {\em Mis-$y$} the mismatch rate is consistently over-estimated by about 10\% , and in the setting {\em Mis-ind} the mismatch
rate is slightly under-estimated. This would be expected since substantial correlations within blocks of observations reduce the impact of mismatch error.   

\subsection{Logistic GLM}
We assume a logistic regression model for the outcome variable given one
binary covariate $d$ (taking the values zero and one in equal proportions), one continuous covariate $x$ (uniformly spaced between $-3$ and $3$), and the associated interaction term. In short, we have 
\begin{equation*}
\text{logit}(\p(y = 1|x, d)) = \beta_0^* + \beta_1^* d + \beta_2^* x + \beta_3^* (x \cdot d),
\end{equation*}
where $\beta_0^* = 0.5$, $\beta_1^* = -1.5$, $\beta_2^* = 1$, and $\beta_3^* = 0.5$. Regarding the mismatch indicator, we confine ourselves to the scenario {\em Constant} as used in the previous subsection; given a binary
outcome variable and the limited impact of mismatch error in this case, estimating the mismatch rate yields a significant challenge even when it is constant. 

Given a binary outcome, $f_{\M{y}}$ is estimated based on the empirical frequencies of $\{ y_i = 1\}_{i = 1}^n$
and $\{ y_i = 0 \}_{i = 1}^n$. The mismatch rate is estimated on the logit scale (cf.~Table \ref{tab_sim_logistic}). We conduct 10k replications.    

{\em Results}. While Table \ref{tab_sim_logistic} confirms that the proposed approach has moderate merits as the mismatch rate increases, the comparison to a plain GLM fit ignoring mismatches is rather far from being as clear as in the Poisson case. For mismatch rates of 5\% or less, the plain GLM fit has comparable bias, less standard deviation, and still roughly maintains the nominal coverage rate. While the proposed approach keeps the relative bias roughly constant, standard deviations increase by factors of up to $2.5$ as the mismatch rate reaches 30\%. Accordingly, confidence intervals tend to be wider and coverage rates consistently exceed the nominal level, whereas the plain GLM fit yields dramatic under-coverage as the mismatch rate is increased. Estimation of the logit of $1 - \alpha^*$ (i.e., the correct match rate) is associated with an upward bias and significant variation, which is unsurprising given that about half of the mismatches will preserve the original value of the response. In summary, the results thus confirm that as the response carries less information (with a binary response being an extreme case), adjustment for mismatch error and estimation of the underlying parameters becomes a much more challenging task.      
 \begin{table}
\begin{center}
\footnotesize
\begin{tabular}{|l|l|ccc|ccc|ccc|ccc|ccc|}
\hline
\multicolumn{1}{|c|}{}& \multicolumn{1}{|c|}{} & \multicolumn{3}{c|}{$\wh\beta_0$}&  \multicolumn{3}{c|}{$\wh\beta_1$}& \multicolumn{3}{c|}{$\wh\beta_2$} & \multicolumn{3}{|c|}{$\wh\beta_3$} & \multicolumn{3}{c|}{$\wh\gamma$} \\ \hline
$\alpha^*$& & RB & SD & CG &   RB & SD & CG & RB & SD & CG & RB & SD & CG & RB & SD & CG  \\ \hline 
     0    & {\bfseries \textsf{w/}} & $.10$ & $.21$ & $.97$ & $.11$ & $.35$ & $.98$ & $.08$ & $.18$ & $.98$ & $.17$ & $.26$ & $.97$ & & & \\
        & {\bfseries \textsf{w/o}}   & .01 & $.19$ & $.95$ & $.01$ & $.28$ & $.95$ & $.01$ & $.13$ & $.95$ & $.01$ &  $.21$ & $.95$ &  & &  \\
             \hline
                .05   &  {\bfseries \textsf{w/}}   & $.08$ & $.23$ & $.98$ & $.09$ & $.41$ & $.99$ & $.07$ & $.22$ & $.98$ & $.13$ & $.30$ & $.97$ & $.18$ & $1.8$ & $.95$ \\
             & {\bfseries \textsf{w/o}}  & $-.08$ & $.19$ & $.95$  & $-.10$ & $.28$ & $.92$ & $-.08$  & $.13$ & $.91$ & $-.14$ & $.20$ & $.94$ & & & \\
\hline          
    .10     & {\bfseries \textsf{w/}}   & $.08$ & $.26$ & $.98$ & $.09$ & $.48$ & $.98$ & $.07$ & $.27$ & $.98$ & $.13$ & $.34$ & $.97$ & $.21$ & $1.5$ & $.98$  \\
            & {\bfseries \textsf{w/o}}       &  $-.17$ & $.19$ & $.93$ & $-.18$ & $.28$ &  $.83$  & $-.15$ & $.13$ &  $.77$ & $-.25$ & $.19$ & $.90$ & & & \\
              \hline          
      .15   & {\bfseries \textsf{w/}}   & $.08$ & $.28$ & $.98$  & $.08$ & $.53$ & $.98$ & $.06$ & $.30$ & $.98$ & $.13$ & $.39$ & $.97$ & $.21$ & $1.3$ & $.99$ \\
           & {\bfseries \textsf{w/o}}   &  $-.25$ & $.19$ & $.90$ & $-.26$ & $.27$ & $.68$ & $-.22$ & $.12$ & $.54$ & $-.34$ & $.18$ & $.84$ & & & \\
          \hline
  .20     & {\bfseries \textsf{w/}}  & $.08$  & $.30$  & $.97$  & $.09$ & $.59$ & $.98$ & $.06$ & $.34$  & $.97$ & $.14$ & $.44$ & $.97$ & $.23$ & $1.1$ & $.99$ \\
           & {\bfseries \textsf{w/o}}   & $-.31$ & $.19$ & $.87$ & $-.34$ & $.27$ & $.53$ & $-.29$ & $.12$  & $.32$ & $-.42$ & $.18$ & $.78$ & & & \\
                       \hline
    .25  &  {\bfseries \textsf{w/}}  & $.10$  & $.34$  & $.97$  &  $.11$ & $.66$ & $.98$ & $.08$ & $.40$  & $.97$ & $.15$ & $.49$ & $.96$ & $.22$ & $.95$ & $.99$  \\
          & {\bfseries \textsf{w/o}}    & $-.38$ & $.19$ & $.83$ & $-.40$ & $.27$ & $.38$  & $-.35$ & $.12$ & $.16$ & $-.49$  & $.17$ & $.69$ &  &  & \\
                           \hline
    .30   & {\bfseries \textsf{w/}}   & $.12$ & $.38$ & $.97$ & $.12$ & $.75$ & $.99$  & $.09$ & $.47$  & $.97$ & $.16$ & $.61$ & $.96$ & $.28$ & $.90$ & $.99$ \\
          & {\bfseries \textsf{w/o}}   & $-.42$ & $.19$ & $.80$ & $-.45$ & $.27$ & $.27$ & $-.40$ & $.12$ & $.07$ & $-.55$ & $.17$ & $.62$ &  & & \\
             \hline
\end{tabular}
\end{center}
\caption{Results of the simulation study for the logistic GLM based on 10k replications. The mismatch rate is given by 
$\alpha^*$. Here, we estimate the parameter $\gamma^* = \log((1- \alpha^*)/\alpha^*)$ (rightmost column). RB -- relative bias; SD -- standard deviation; CG -- coverage rate of confidence intervals. {\bfseries \textsf{w/}}: with adjustment (proposed approach); {\bfseries \textsf{w/o}}: without adjustment, i.e., the naive estimator (plain GLM) ignoring mismatch error.}\label{tab_sim_logistic}
\end{table}

\section{Applications}\label{sec:apps}

In this section, we illustrate our methodology in three case studies involving real data sets obtained from
record linkage including (i) a longevity analysis based on historical linkage, (ii) analysis of two-way
contingency tables obtained from linking Medicare claims and survey responses, and (iii) the investigation
of time trends in the issuance of nurse licenses. The data sets for (i) and (iii) are open access\footnote{(i) \url{openicpsr.org/openicpsr/project/155186/}\\ \hspace*{4ex}(iii) \url{data.wa.gov/Health/Health-Care-Provider-Credential-Data/qxh8-f4bd}}.

\subsection{Longevity analysis}\label{subsec:mp}
The Life-M project (see \url{life-m.org} for details) provides multi-generational data from the 20th century that was gathered from various data sources including birth certificates, death certificates, marriage certificates, and decennial censuses. In our case study, we study the relationship between age of death and year of birth obtained from linking birth and death certificates. LIFE-M used a hybrid of two linkage procedures: a fraction of the records were selected for manual linkage by trained research assistants (``hand-linked" records); the remaining records were linked based on probabilistic record linkage without clerical review (``machine-linked" records). The latter records are more inclined to have mismatch errors \citep{LifeMcit}.  

Initial analyses of these data suggest that the death record sources and collection periods influence the trend in the age at death as a function of the year of birth. Therefore, we focus on birth cohorts where longevity tends to increase overall as expected. After visual exploration of the entire data available ($n \approx 155\text{k}$), we decided to use a cubic polynomial to model the relationship between year of birth and age at death (dependent variable); a cubic fit is used to capture the non-linear relationship between the two variables during a specific time period (1883--1906).

Our approach is applied as follows. We assume a Gaussian regression model for the predictor-response relationship. 
Regarding the latent mismatch indicators $\{ m_i \}_{i = 1}^n$, we assume that all $2,159$ hand linked records are correctly matched (i.e., for the corresponding records it holds that $m_i = 0$). For machine-linked records the mismatch indicator is considered unknown, and is modeled via a logistic regression model whose predictors 
are given by commonness of the first name (\verb+commf+) and the last name (\verb+comml+) of the associated individuals. Since these variables are readily available and probabilistic record linkage was primarily based on names, they are considered suitable surrogates in lieu of more specific information about the correctness of matches as would be output by a probabilistic record linkage procedure. The marginal distribution of the response variable is assumed to follow a Gaussian distribution whose parameters are estimated from the entire data and subsequently treated as fixed. In summary, inference is based on the following specifications:
\begin{align}\label{eq:approach_lifem}
\begin{split}
&y_i|x_i, \{ m_i = 0 \} \sim N(\beta_0 + \beta_1 x_i + \beta_2 x_i^2 + \beta_3 x_i^3, \sigma^2), \qquad 
y_i| \{ m_i = 1 \} \sim N(\mu, \tau^2), \\[.7ex] 
&m_i|\texttt{commf}_i, \, \texttt{comml}_i \sim \text{Bernoulli} \left(\frac{\exp(\gamma_0 + \gamma_1 \cdot \texttt{commf}_i + \gamma_2 \cdot \texttt{comml}_i)}{1 + \exp(\gamma_0 + \gamma_1 \cdot \texttt{commf}_i + \gamma_2 \cdot \texttt{comml}_i)} \right), \; \; 1 \leq i \leq n.  
\end{split}
\end{align}
Additionally, the Life-M team expects the mismatch rate among the machine-linked rates to be around 5\% \citep{LifeMcit}. This information is incorporated by imposing corresponding constraints on the average 
of the linear predictors as described in $\S$\ref{subsec:mismatch_modeling}. Specifically, we consider
the upper bounds $-3$ and $-2.5$ on the logit scale, corresponding to about 5\% and 7.5\%, respectively,
on the probability scale; the latter bound allows for a slightly higher fraction of mismatches as expected. 

\begin{figure}
\begin{center}
\includegraphics[height = 0.52\textwidth]{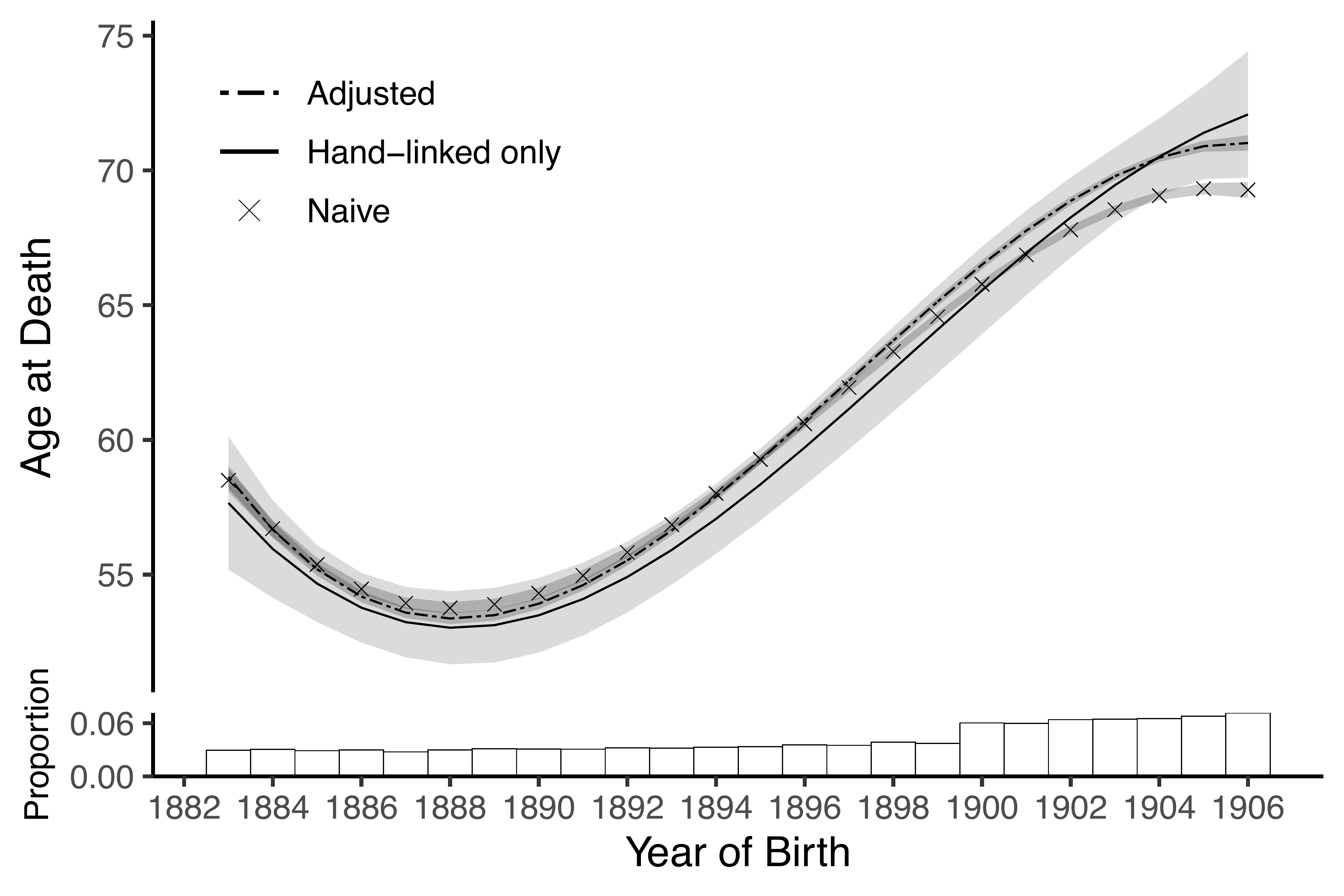}
\end{center}
\vspace*{-4ex}
\caption{Predicted survival times for birth cohorts 1883--1906 based on the Life-M data with point-wise confidence intervals (grey-shaded areas; the wide light grey area corresponds to the results based on the ``hand-linked" records only). "Adjusted" refers to the results under model \eqref{eq:approach_lifem} with the average linear predictor in the model for the $\{ m_i \}$ upper bounded by $-3$ ($\sim$5\% mismatch rate). The proportion of the birth cohorts is shown at the bottom of the plot.}\label{fig:LIFEM}
\end{figure}

Results are summarized in Figure \ref{fig:LIFEM} and Table \ref{tab:LIFEM}. The estimated coefficients and predictions of the cubic fit generated by approach \eqref{eq:approach_lifem} are well within the realm of 
the naive analysis without adjustment for mismatches and an analysis confined to the much smaller subset of
hand-linked records only. Figure \ref{fig:LIFEM} indicates that predictions under the naive and adjusted approaches
start diverging from birth cohort 1897, with predictions under the naive approach falling below those under the 
adjusted approach and those under the hand-linked only analysis. Adjustment yields small reductions of the estimated
residual standard error (about 2.5\% and 4\%, respectively) and the standard errors of the coefficients of the cubic polynomial tend to be slightly smaller as well. First name commonness and last name commonness are both predictive of the latent match status. The sign of the coefficient for first name commonness is unexpected though (since intuitively the more common a name, the more likely mismatches tend to occur). For this reason, we also explored
the use of an interaction model with the same two predictor variables but since this change neither improved interpretability nor model fit we decided to retain the main effect model.

\begin{table}
\centering
\begin{tabular}{|l|cc|cc|cc|cc|cc|}
\hline
 & \multicolumn{2}{|c|}{Naive}& \multicolumn{2}{|c|}{Adjusted$^{\ddagger}$} & \multicolumn{2}{|c|}{Adjusted$^{\dagger}$}  & \multicolumn{2}{|c|}{Hand-linked} \\ \hline
$\wh\beta_0$ & $58.5$  & $(0.2)$ & $58.6$ & $(0.1)$ & $58.7$ & $(0.2)$ & $57.7$ & $(1.26)$  \\ \hline
$\wh\beta_1$  & $-46.7$ & $(1.8)$ & $-51.0$ & $(1.5)$ & $-52.5$ & $(1.6)$ & $-44.2$ & $(11.6)$ \\ \hline
$\wh\beta_2$ & $130.4$ & $(4.0)$ & $140.2$ & $(3.9)$ & $143.2$ & $(3.9)$ & $118.6$ &  $(27.9)$  \\ \hline
$\wh\beta_3$ & $-72.9$ & $(2.5)$ & $-76.8$ & $(2.6)$ & $-77.7$ & $(2.7)$ &  $-59.9$ & $(18.5)$  \\ \hline
$\wh\sigma$ & $21.2$ & (0.04) &  $20.7$ &  (0.06) &  $20.4$ & (0.06) & $19.0$  & (0.29) \\ \hline \hline
$\wh\gamma_0$ &  & & $-5.98$ & $(.50)$& $-4.86$ & $(.38)$ & & \\
\hline
$\wh\gamma_1$ & & &  $-1.45$ &$(.55)$ & $-1.38$ & $(.43)$ & & \\
\hline
$\wh\gamma_2$ & & &  $7.2$ & $(.33)$  & $6.05$ & $(.25)$  & & \\
\hline
\end{tabular}\label{tab:LIFEM}
\caption{Parameter estimates for the longevity analysis (standard error in parentheses). The symbols $\dagger$ and $\ddagger$ refer to the bounds $-3.0$ and $-2.5$ on the average of the linear predictors in the model for the mismatch indicators (cf.~$\S$\ref{subsec:mismatch_modeling})}
\end{table}

\subsection{Agreement of Medicare claims and survey responses}\label{subsec:hsr_medicare}

The second case study presents an application of the two-way contingency table methodology outlined in Subsection \ref{subsec:up}. This case study is based on a linkage between a survey conducted as part of the Health and Retirement Study (HRS, see \url{hrs.isr.umich.edu}), the largest and most comprehensive nationally representative multi-disciplinary panel study of Americans over the age 50, and Medicare claims data. Such linkages to administrative data are routinely performed by HRS researchers, e.g., to supplement or validate data reported by survey respondents given their consent to link.


In 2020, the HRS re-evaluated the Medicare record linkage performed in 2018, and identified 59 cases in the 2018 linkage that were likely mismatches, either because these cases were linked to \emph{different} claims records in the new linkage in 2020, or these cases could not be linked to a claims record in 2020. In this case study, we focus on the effects of including these likely mismatched cases in a contingency table analysis of the 2018 HRS data. Specifically, we look at the bivariate association between self-reports of nursing home attendance in the past two years and administrative records of nursing home attendance in that same time frame. Of specific interest to the HRS is the level of agreement between these two measures, for the purpose of investigating potential measurement error. 

We note that the overall rate of likely mismatches in 2018 is rather small, given that there were 8,665 consenting respondents in total. For the sake of this illustration, we therefore selected a simple random sample of 300 HRS respondents who were not deemed to be mismatches in 2018, effectively simulating a mismatch rate of 59/359 = 0.164. In this scenario, the mismatched cases may have an effect on the contingency table analysis.

In the analysis, we used the four proportions defining the two-by-two contingency table based on the 300 exact matches as the benchmark proportions for evaluation. We computed the mean relative absolute error (MRAE) of the four proportions defining the contingency table, the Kullback-Leibler divergence (KLD) as a measure of distance between the proportions in the contingency table and the proportions based on the exact matches, the one-sample chi-square goodness of fit measure for the four proportions (again using the benchmark proportions), the chi-square measure of association between the two variables, and Cohen's kappa statistic.

Table \ref{tab:HRS} presents the results of our analysis. Compared to an analysis of the 300 known exact matches, the naive analysis of the 359 cases would result in a higher MRAE of the four proportions defining the table, a larger KLD based on the four proportions, a larger chi-square goodness of fit (GOF) measure, an attenuated chi-square measure of association, and an attenuated kappa statistic (understating the level of agreement between the two variables). The adjustment approach described in Subsection \ref{subsec:up} assuming the aforementioned mismatch rate of $0.164$ would reduce the errors and yield measures of agreement that are more consistent with the known true values.

\begin{table}
\centering
\begin{tabular}{|l|l|l|l|l|l|}
\hline
 & MRAE & KLD & GOF & Association & Kappa \\ \hline
Naive & 0.1685 & 0.0020 & 1.4780 & 125.12 & 0.6130 \\ \hline
Adjusted & 0.1334 & 0.0017 & 1.0995 & 147.06 & 0.6625 \\ \hline
Exact & 0.0000 & 0.0000 & 0.0000 & 146.80 & 0.6569 \\ \hline
\end{tabular}
\caption{Results of the HRS-Medicare claims contingency table analysis. ``Exact" refers to the analysis based on the correct matches only and is used as benchmark.}\label{tab:HRS}
\end{table} 

\subsection{Investigation of trends in nurse license processing times}\label{subsec:hcc}
In this section, we evaluate the utility of the proposed approach on curve fitting via penalized splines (cf.~$\S$\ref{subsec:Bayesian}). Specifically, we study an application to a nurse credential database from the state of Washington\protect\footnote{https://data.wa.gov/Health/Health-Care-Provider-Credential-Data/qxh8-f4bd} between 01/01/2009 and 12/31/2021.\\[1ex] 
\noindent {\bfseries Data and Linkage}.
Each entry in this database corresponds to one specific nurse practice license issued to one specific nurse, containing the following information: full name of the nurse and their year of birth, credential number, issue and expiration dates, status (active, closed, expired), and type of license (e.g., ``registered nurse license", "medical assistant certification", "registered nurse temporary practice permit"). Nurses are commonly issued a temporary permit 
prior to receiving a regular license. In our study, we investigate the average duration of the associated transitional period (in \#days), which is of interest to researchers in health metrics \citep{Flaxman22}.  For this purpose, the two data subsets corresponding to temporary permits and regular licenses, respectively, are extracted and subsequently linked. 

Data linkage is performed by first blocking on year of birth and first initial of the last name of the nurse, and then string matching of first, middle, and last names within each block using the Jaro-Winkler metric \citep{JaroWinkler1990}. We consider both exact name matching (``restrictive linkage") and inexact name matching (``generous linkage"); in the latter case, two records are declared a match as long as the Jaro-Winkler match scores for each name variable exceeds the threshold .85 (chosen ad-hoc via visual inspection of the histograms of the scores). The resulting restrictively linked and generously linked files consist of about 61k and 78k records, respectively, after removing records that were obvious mismatches since the associated lengths of the waiting period between permits were negative. 

Ranges for the underlying mismatch rates in these two files were determined as follows: the first estimate assumes that the number of mismatches is about the same as the number of obvious mismatches associated with negative durations; the second estimate is based on excessively large durations ($\geq 1.5$ years). This yields the range 
$[3.7\%, 8.1\%]$ for the generously linked file and $[0.4\%,  1.0\%]$ for the restrictively linked file. The still noticeable fraction in the latter file despite exact name matching can be attributed mostly to multiple instances of the license issue process for the same nurse. Given the available information, it is unclear how to determine 
the true match status with certainty even with a clerical review: in case of multiple issuances, only the earliest and latest dates of issuance are recorded, i.e., any intermediate dates are not given.  \\[1ex]
\noindent {\bfseries Post-Linkage Analysis}. The goal of the analysis is to identify trends/variations over time in the average duration of the aforementioned transitional period from the time a temporary permit is issued until it gets substituted by a regular nurse license. We let $\{ x_i \}_{i = 1}^n$ denote the temporary permit issue dates (scaled to $[0,1]$ such that 
$01/01/2009$ and $12/31/2021$ correspond to $0$ and $1$, respectively) and consider the duration until the regular license issue date as the dependent variables $\{ y_i \}_{i = 1}^n$. The latter are obtained from the linked files and hence in part incorrect as a consequence of mismatch error (since observations with negative durations are dropped, the remaining mismatch error tends to produce inflated durations). In order to flexibly capture trends in average duration over time, the corresponding mean function for correctly matched observations is modeled via a cubic spline, i.e., 
\begin{equation*}
\E[y_i|x_i, m_i = 0;\bm\beta] = s_{\bm\beta}(x_i), \quad 1 \leq i \leq n,  \quad s_{\bm\beta}(x) = \sum_{j = 1}^d \beta_j B_j(x),
\end{equation*}
where the $\{ B_j \}_{j = 1}^d$ represent the associated B-spline basis functions given 1,000 knots placed evenly in
$[0,1]$. Conditional on $\{ m_i = 1 \}$, we assume an intercept-only model for the dependent variable $y_i$, $1 \leq i \leq n$. For simplicity, we assume Gaussian models (with the different variances) for each of these specifications in order to apply the proposed approach, but alternative models (e.g., Poisson) could be used as well. 

\begin{figure}
\begin{minipage}{.55\textwidth}
\hspace*{6ex}\includegraphics[width = \textwidth]{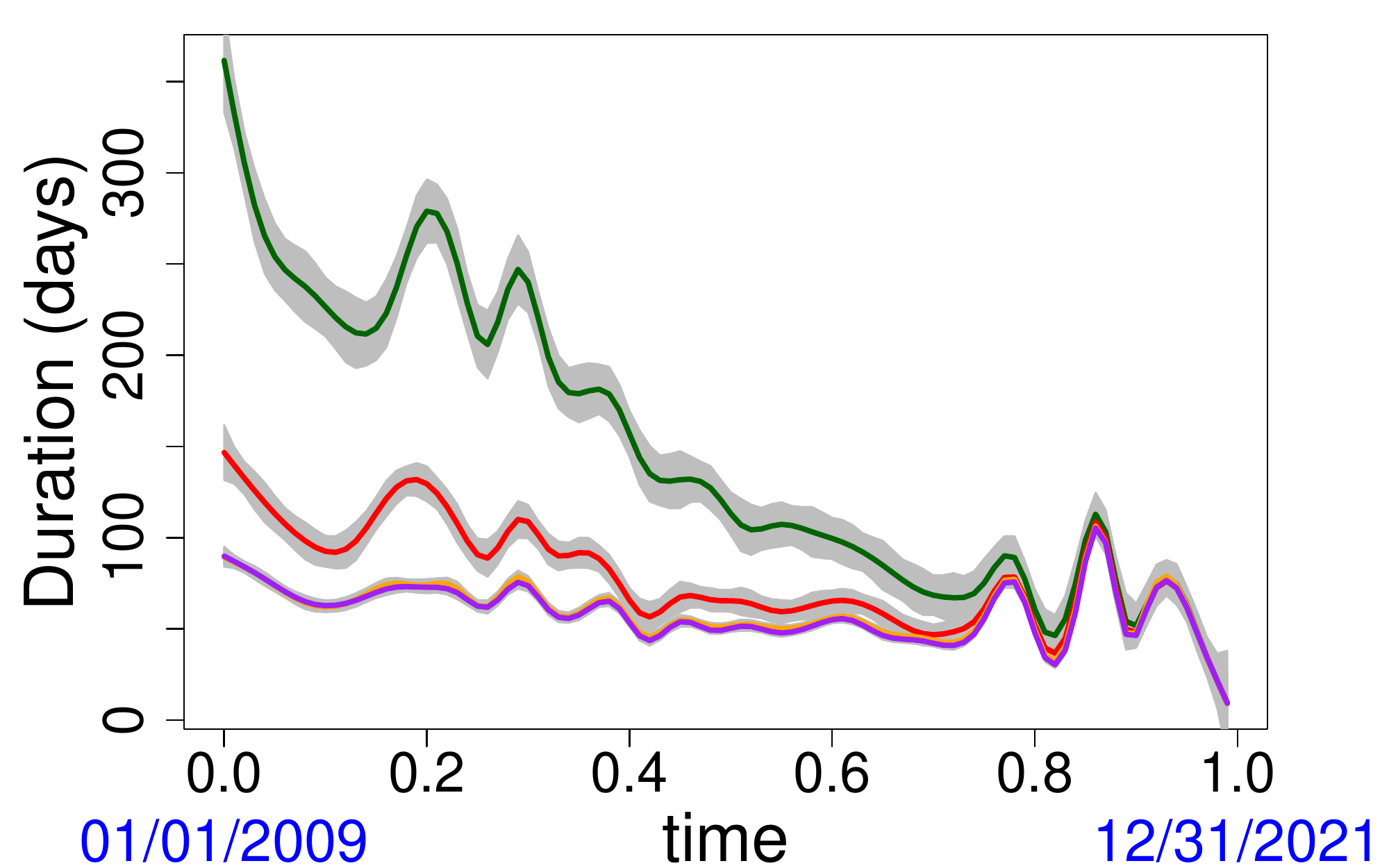}
\end{minipage}
\begin{minipage}{.43\textwidth}
\vspace*{-3.5ex}
 { \footnotesize
  \begin{center}
  \begin{tabular}{l}    
  \textcolor{gmu}{\rule[.5ex]{4ex}{.3ex}} generously
  linked, ``naive'' \\[1ex]
  \textcolor{red}{\rule[.5ex]{4ex}{.3ex}} restrictively linked,
  ``naive'' \\[1ex]
  \textcolor{orange}{\rule[.5ex]{4ex}{.3ex}} generously linked,
  adjusted \\[.33ex]
  $\wh{\alpha}_{\text{gen}} = 7.2\% \;\;[6.95\%, \, 7.47\%]$ \\[1ex]
  \textcolor{violet}{\rule[.5ex]{4ex}{.3ex}} restrictively linked,
  adjusted \\[.33ex]
    $\wh{\alpha}_{\text{res}} = 4.70\% \;\;[4.30\%, \, 5.00\%]$
  \end{tabular}  
 \end{center}
}
\end{minipage}
\caption{{\small Estimated mean functions for the duration of regular nurse license issuance with and without adjustment for mismatch error based on the generously and restrictively linked files. As explained in the text, for the results after adjustment we report results for which the mismatch rate hits a plateau among the range of solutions under consideration as given in Table \ref{tab:credential}.}}\label{fig:credential}
\end{figure}

The Bayesian inference approach outlined in 
$\S$\ref{subsec:Bayesian} is applied to both the generously and the restrictively linked data set. The number of MCMC iterations is set to 10,000 after a burn-in period of length 100, out of which every tenth MCMC sample is retained for posterior inference. In addition to an ``out-of-the-box" application, we also run the approach with the residual
standard deviation $\sigma$ of the spline regression model fixed to a range of fractions $\{.1,.15,\ldots,.95,1\}$ of the residual standard deviation $\wh\sigma_0$ from
a ``naive" spline fit without accounting for mismatches. While the resulting mean functions do not change substantially, the additional (varying) constraint on $\sigma$ allows us to explore a range of plausible solutions and associated estimates of the mismatch rate. The ratio $\sigma / \wh{\sigma}_0$ can be interpreted as the relative reduction in root mean squared error after accounting for mismatch error. Figure \ref{fig:credential} shows that without adjustment for mismatches, the estimated mean functions fluctuate strongly at the beginning of the time line; even when the 
restrictively linked file is used, the average duration exhibits fluctuations of $\sim$50 days. The impact of mismatch error
is indeed expected to be more pronounced at the beginning of the time period than towards the end since excess durations 
resulting from incorrect linkage can be more drastic. Interestingly, a second window of rapid fluctuations is observed between
$.75$ and $.85$ (scaled time scale), however, these fluctuations are present before and after adjustment for mismatch error and are hence more likely to be genuine. Moreover, after adjustment the estimated mean functions are significantly closer to the estimated mean functions (unadjusted) based on the restrictively linked file, and the estimated mean functions after adjustment are essentially identical {\em regardless of whether adjustment was based on the the generously or the restrictively linked file}. Table \ref{tab:credential} shows that for the generously linked file, the estimated mismatch rate plateaus for $\wh{\sigma} / \wh{\sigma}_0 = 0.4$ yielding an estimate of 7.2\% of mismatches, which is well within the anticipated range
between 3.7\% and 8.1\%. For the restrictively linked file, the estimated mismatch rate plateaus for $\wh{\sigma} / \wh{\sigma}_0 = 0.65$ at the value 4.7\%, which is still within the realm of the anticipated range and significantly lower than the estimate
based on the generously linked file. 


\begin{table}
\begin{center}
{\footnotesize \begin{tabular}{|l|lllllllllll|}
\hline & & & & & & & & & & & \\[-3ex]
$\wh{\sigma}_{\text{gen}} / \wh{\sigma}_0$ & $.08^{\star}$ & $0.1$& $0.2$& $0.3$& $0.4$& $0.5$ & $0.6$& $0.7$ & $0.8$ & $0.9$ & $1$ \\
\hline
$\wh{\alpha}_{\text{gen}}$ & $.161$  & $.135$ & $.081$  & $.073$ & $.072$ & $.076$ & $.084$ & $.098$ &  $.124$ & $.175$ & $.96$\\
& {\scriptsize $[.158,$} & {\scriptsize $[.132,$}  & {\scriptsize $[.079,$}  & {\scriptsize $[.070,$} & {\scriptsize $[.070,$} & {\scriptsize $[.073,$} &  {\scriptsize $[.081,$} &  {\scriptsize $[.095,$} & {\scriptsize $[.118,$} & {\scriptsize $[.166,$} &  {\scriptsize $[.961,$} \\
& {\scriptsize $.164]$} & {\scriptsize $.138]$}  & {\scriptsize $.084]$}  & {\scriptsize $.075]$} & {\scriptsize $.075]$} & {\scriptsize $.078]$} &  {\scriptsize $.088]$} &  {\scriptsize $.103]$} & {\scriptsize $.130]$} & {\scriptsize $.184]$} &  {\scriptsize $.965]$}
 \\
\hline & & & & & & & & & & & \\[-3.5ex]
\hline & & & & & & & & & & & \\[-3ex]
$\wh{\sigma}_{\text{res}} / \wh{\sigma}_0$ & $.14^{\star}$ & $0.1$& $0.2$& $0.3$& $0.4$& $0.5$ & $0.6$& $0.7$ & $0.8$ & $0.9$ & $1$  \\
\hline
$\wh{\alpha}_{\text{res}}$ & $.188$ & $.274$ & $.132$ & $.081$ & $.058$ &  $.049$ & $.046$ & $.047$ & $.053$ & $.073$ &  $.051$ \\
& {\scriptsize $[.183,$} & {\scriptsize $[.269,$}  & {\scriptsize $[.128,$}  & {\scriptsize $[.078,$} & {\scriptsize $[.055,$} & {\scriptsize $[.047,$} &  {\scriptsize $[.043,$} &  {\scriptsize $[.043,$} & {\scriptsize $[.049,$} & {\scriptsize $[.066,$} &  {\scriptsize $[.017,$} \\
& {\scriptsize $.193]$} & {\scriptsize $.278]$}  & {\scriptsize $.135]$}  & {\scriptsize $.084]$} & {\scriptsize $.061]$} & {\scriptsize $.052]$} &  {\scriptsize $.049]$} &  {\scriptsize $.050]$} & {\scriptsize $.058]$} & {\scriptsize $.081]$} &  {\scriptsize $.144]$} \\
\hline
\end{tabular}}
\end{center}
\vspace*{-2.5ex}
\caption{{ \small Estimated mismatch rates (posterior means) and 95\% credible intervals (bracketed, small font) for the nurse credential data depending on the ratio of residual standard deviation $\wh{\sigma}/\wh{\sigma}_0$ before and after adjustment; in the leftmost column (asterisked) $\wh{\sigma}$ is unrestricted, while for the remaining columns the ratio $\wh{\sigma}/\wh{\sigma}_0$ is fixed to the value in the column header. The top half of the table contains the results for the generously linked data set (``gen"), and the bottom half to the restrictively linked data set (``res").}}\label{tab:credential}
\end{table}

\section{Conclusion}\label{sec:conc}
In this paper, we have developed a general framework to enable valid post-linkage inference in the presence of mismatch error in the challenging secondary analysis setting. The proposed framework is flexible in the sense that limited information about the linkage process can be incorporated, and that the same machinery can be applied to handle various models for the mismatch indicator and the linked substantive variables. The approach is scalable and convenient from the perspective of implementation. Results from simulations and case studies with real data consolidate the usefulness for post-linkage analysis.  

At the same time, the work presented here prompts various avenues of future research. First, it is of interest to further investigate the sensitivity of our approach vis-\`a-vis violations of the main assumptions even though the simulations shown here indicate at least a moderate degree of robustness. Second, it is worthwhile to consider extensions covering linkage of more than two files. Third, while mismatch error has undoubtedly received much more attention, false non-matches (missed matches) are similarly important; handling both types of error in an integrated fashion is a desirable goal. Finally, our approach for contingency table analysis highlights a connection to synthetic data methods such a post-randomization \citep{Gouweleeuw1998} for disclosure control, and it would appear to be worth elaborating on that connection in more detail.  
\vskip2ex
\noindent {\bfseries Acknowledgments}. We would like to thank Jessica Faul for providing the data used in $\S$\ref{subsec:hsr_medicare} and Abraham Flaxman for suggestions and discussions leading to the analysis in 
$\S$\ref{subsec:hcc}. 



\bigskip






\bibliographystyle{abbrvnat}
\bibliography{references_M}

\appendix

\section{Standard error calculations}
We here provide specific expressions for \eqref{eq:score_hessian} in regression setups in which both the regression model $\phi(y_i | \M{x}_i;\bm{\theta}) = \phi(y_i | \M{x}_i^{\T} \bm{\beta})$ and the model for mismatch $h(\M{z}_i; \bm{\gamma}) = h(\M{z}_i^{\T} \bm{\gamma})$ are functions of linear predictors, $1 \leq i \leq n$. This covers a wide range of scenarios of practical interest; for linear regression with unknown scale parameter, we refer to \cite{SlawskiDiaoBenDavid2019}. In this setting, we obtain the following expressions
\begin{align}\label{eq:weights_score}
\begin{split}
\nabla_{\bm{\theta}} \, \ell_i(\bm{\theta}, \bm{\gamma}) = \nabla_{\bm{\beta}} \, \ell_i(\bm{\beta}, \bm{\gamma}) &= -\frac{h(\M{z}_i^{\T}\bm{\gamma}) \cdot \phi'(\M{y}_i|\M{x}_i^{\T} \bm{\beta}) \, 
  }{f_{y}(\M{y}_i)  \cdot (1 - h(\M{z}_i^{\T}\bm{\gamma})) + h(\M{z}_i^{\T}\bm{\gamma})
\cdot \phi(\M{y}_i
| \M{x}_i^{\T}\bm{\beta}) } \, \M{x}_i, \\[1.5ex]
\nabla_{\bm{\gamma}} \, \ell_i(\bm{\beta}, \bm{\gamma})  &= -\frac{
                                    h'(\M{z}_i^{\T}\bm{\gamma})
                                    (\phi(\M{y}_i|\M{x}_i^{\T}
                                    \bm{\beta}) - f_{\M{y}}(\M{y}_i))}{f_{y}(\M{y}_i)  \cdot \big(1 - h(\M{z}_i^{\T}\bm{\gamma}) \big) + h(\M{z}_i^{\T}\bm{\gamma}) 
\cdot \phi(\M{y}_i
| \M{x}_i; \bm{\theta}) } \, \M{z}_i,
\end{split}
\end{align}
where we have used the shortcuts 
\begin{equation}\label{eq:shortcuts_derivatives}
\phi'(\M{y}_i|\eta) \coloneq \frac{d}{d\eta} \phi(\M{y}_i|\eta) \dev{\eta}{\M{x}_i^{\T} \bm{\beta}}, \qquad 
h'(\M{z}_i^{\T}\bm{\gamma}) \coloneq \frac{d}{d\zeta} h(\zeta) \dev{\zeta}{\M{z}_i^{\T} \bm{\gamma}}, \quad 1 \leq i \leq n. 
\end{equation}
This yields 
\begin{equation*}
\su \nabla \ell_i(\bm{\beta}, \bm{\gamma})^{\otimes2} = \begin{bmatrix}
                             \M{X}^{\T} \M{W}_1^2 \M{X} &  \M{X}^{\T}
                             \M{W}_1 \M{W}_2 \M{Z} \\[1ex]
                             \M{Z}^{\T} \M{W}_2 \M{W}_1 \M{X} &  \M{Z}^{\T}
                             \M{W}_2^2 \M{Z}
                            \end{bmatrix}, \qquad 
\end{equation*}
where $\M{W}_1$ and $\M{W}_2$ are diagonal matrices whose diagonal entries are given by the expressions in \eqref{eq:weights_score} preceding $\M{x}_i$ and $\M{z}_i$, respectively, $1 \leq i \leq n$, and $\M{X}$ and $\M{Z}$ are the matrices whose
rows are given by $\{ \M{x}_i \}_{i = 1}^n$ and $\{ \M{z}_i \}_{i = 1}^n$. 

Next, we calculate 
\begin{align}\label{eq:weights_hessian}
\begin{split}
\nabla_{\bm{\beta}}^2  \, \ell_i(\bm{\beta}, \bm{\gamma}) = -\frac{h(\M{z}_i^{\T} \bm{\gamma})
\phi''(\M{y}_i | \M{x}_i^{\T} \bm{\beta})}{f_{y}(\M{y}_i)  \cdot \big(1 - h(\M{z}_i^{\T}\bm{\gamma}) \big) + h(\M{z}_i^{\T}\bm{\gamma})
\cdot \phi(\M{y}_i
| \M{x}_i^{\T} \bm{\beta})} \M{x}_i \M{x}_i^{\T} \\[1ex] + \frac{h(\M{z}_i^{\T} \bm{\gamma})^2 \left( \phi'(\M{y}_i|\M{x}_i^{\T} \bm{\beta}) \right)^2}{\left[ f_{y}(\M{y}_i)  \cdot \big(1 - h(\M{z}_i^{\T}\bm{\gamma}) \big) + h(\M{z}_i^{\T}\bm{\gamma})
  \cdot \phi(\M{y}_i| \M{x}_i^{\T} \bm{\beta}) \right]^2} \M{x}_i \M{x}_i^{\T} \\[2ex]
\nabla_{\bm{\gamma}}^2  \, \ell_i(\bm{\beta}, \bm{\gamma}) =     -\frac{
                                    h''(\M{z}_i^{\T}\bm{\gamma})
                                    \big(\phi(\M{y}_i|\M{x}_i^{\T} \bm{\beta}) - f_{\M{y}}(\M{y}_i) \big)}{f_{y}(\M{y}_i)  \cdot \big(1 - h(\M{z}_i^{\T} \bm{\gamma}) \big) 
                                    + h(\M{z}_i^{\T}\bm{\gamma})
\cdot \phi(\M{y}_i
                                                                 |
                                                                 \M{x}_i^{\T}
                                                                 \bm{\beta})
                                                                 } \M{z}_i \M{z}_i^{
                                                                 \T}\\[1ex]
                                                                 +
                                                                 \frac{
                                    (\phi(\M{y}_i|\M{x}_i^{\T} \bm{\beta}) -
                                                                 f_{\M{y}}(\M{y}_i))^2
                                    \left( h'(\M{z}_i^{\T} \bm{\gamma}) \right)^2}{
                                                                 \left[
                                                                 f_{y}(\M{y}_i)  \cdot \big(1 - h(\M{z}_i^{\T} \bm{\gamma}) \big) + h(\M{z}_i^{\T}\bm{\gamma})
\cdot \phi(\M{y}_i
| \M{x}_i^{\T} \bm{\beta}) \right]^2} \M{z}_i \M{z}_i^{\T} \\[2ex]
\nabla_{\bm{\beta}} \nabla \bm{\gamma}  \ell_i(\bm{\beta}, \bm{\gamma}) = -\frac{
  \phi'(\M{y}_i
| \M{x}_i^{\T} \bm{\beta}) \cdot h'(\M{z}_i^{\T} \bm{\gamma})}{f_{y}(\M{y}_i)  \cdot (1 - h(\M{z}_i^{\T}\bm{\gamma})) + h(\M{z}_i^{\T}\bm{\gamma})
\cdot \phi(\M{y}_i
| \M{x}_i^{\T} \bm{\beta})} \M{x}_i \M{z}_i^{\T} \\[1ex] + \frac{h(\M{z}_i^{\T}\bm{\gamma}) \cdot
  \phi'(\M{y}_i
| \M{x}_i^{\T} \bm{\beta}) \big(\phi(\M{y}_i | \M{x}_i^{\T} \bm{\beta}) - f_{y}(\M{y}_i) \big) h'(\M{z}_i^{\T} \bm{\gamma})}{\left[ f_{y}(\M{y}_i)  \cdot (1 - h(\M{z}_i^{\T} \bm{\gamma})) + h(\M{z}_i^{\T}\bm{\gamma})
\cdot \phi(\M{y}_i
| \M{x}_i^{\T} \bm{\beta}) \right]^2} \M{x}_i \M{z}_i^{\T},
\end{split}
\end{align}    
$1 \leq i \leq n$, where the second derivatives $\phi''$ and $h''$ are defined analogously to \eqref{eq:shortcuts_derivatives}. Putting together the above pieces, we obtain that 
\begin{equation*}
\su \nabla^2 \ell_i(\bm{\beta}, \bm{\gamma}) = \begin{bmatrix}  
 \M{X}^{\T} \M{W}_3   \M{X} &  \M{X}^{\T} \M{W}_4   \M{Z} \\[1ex]
 \M{Z}^{\T}  \M{W}_4  \M{X}   & \M{Z}^{\T}  \M{W}_5  \M{Z} 
\end{bmatrix},
\end{equation*}
where $\M{W}_3$ through $\M{W}_5$ are diagonal matrices whose diagonal entries are given by the terms 
associated with $\M{x}_i \M{x}_i^{\T}$, $\M{x}_i \M{z}_i^{\T}$, and $\M{z}_i \M{z}_i^{\T}$ in , respectively, $1 \leq i \leq n$. 

\end{document}